\documentclass[12pt]{article}
\pdfoutput=1
\newlength{\abstractwidth}
\usepackage{color}
\usepackage{graphicx}

\renewcommand{\thefootnote}{\fnsymbol{footnote}}
\renewcommand{\thanks}[1]{\footnote{#1}}
\newcommand{\starttext}{
\setcounter{footnote}{0}
\renewcommand{\thefootnote}{\arabic{footnote}}}

\newcommand{\bea}{\begin{eqnarray}}
\newcommand{\eea}{\end{eqnarray}}
\newcommand{\ee}{\end{equation}}
\newcommand{\be}{\begin{equation}}

\newcommand{\sm}{\smallskip}

\def\cE{{\cal E}}

\def\cN{{\cal N}}

\def\half{ {1\over 2}}
\def\p{\partial}

\def\l({\left(}
\def\r){\right)}

\def\a{\alpha}
\def\b{\beta}

\def\eps{\epsilon}
\def\g{\gamma}

\def\l{\lambda}

\def\g{\gamma}

\def\ti{\tilde}
\def\hE{\hat E}

\def\E{{\cal E}}

\def\no{\nonumber}

\begin{document}
\starttext
\setcounter{footnote}{0}

\begin{flushright}
24 November  2009
\end{flushright}

\bigskip

\begin{center}

{\Large \bf Charged Magnetic Brane Solutions in
AdS$_5$}

\vskip .25cm

{\large \bf and the fate of the third law of thermodynamics\footnote{This work was
supported in part by NSF grant PHY-07-57702.}}

\vskip .6in

{\large \bf Eric D'Hoker and  Per Kraus}

\vskip .2in

{ \sl Department of Physics and Astronomy }\\
{\sl University of California, Los Angeles, CA 90095, USA}\\
{\tt \small dhoker@physics.ucla.edu; pkraus@physics.ucla.edu}

\end{center}

\vskip .2in

\begin{abstract}

\vskip 0.1in

We construct asymptotically AdS$_5$ solutions to 5-dimensional Einstein-Maxwell theory
with Chern-Simons term which are dual to 4-dimensional gauge theories,  
including $\cN=4$ SYM theory, in the presence of
a constant background magnetic field $B$ and a uniform electric charge density $\rho$.
For the solutions corresponding to supersymmetric gauge theories, we find numerically  that a small magnetic field causes a drastic decrease in the entropy at low temperatures.   The near-horizon AdS$_2 \times R^3$ geometry of the purely electrically charged brane thus appears to be unstable under the addition of a small magnetic field.   Based on this observation, we propose a formulation of the third law of thermodynamics (or  Nernst theorem) that can be applied to black holes in the AdS/CFT context.

\sm

We also find interesting behavior for smaller, non-supersymmetric, values of the
Chern-Simons coupling $k$.  For $k=1$ we exhibit exact solutions corresponding to warped AdS$_3$ black holes, and show that these can be connected to asymptotically AdS$_5$ spacetime.
For $k\leq 1$ the entropy appears to go to a finite value at extremality, but the solutions still exhibit a mild singularity at strictly zero temperature.

\sm

 In addition to our numerics,
we carry out a complete perturbative analysis valid to order $B^2$, and find that this
corroborates our numerical results insofar as they overlap.

\end{abstract}

\newpage

\section{Introduction and summary of results}
\setcounter{equation}{0}

The dynamics of gauge theories at finite temperature, charge density and
background electromagnetic fields may be studied at strong coupling via
AdS/CFT dual supergravity solutions. The supergravity approximation is valid for
large $N$ and large `t Hooft coupling, but may be consistently truncated further
to Einstein/Maxwell theory in the bulk when studying these electromagnetic effects. Temperature arises as the  solutions  to this Einstein/Maxwell theory exhibit a horizon,
while charge density and background electromagnetic fields are introduced by imposing boundary conditions on the bulk Maxwell field.
Thermodynamics and transport properties in gauge theories at strong coupling
may then be obtained from suitable black hole or black brane solutions in this
relatively simple Einstein/Maxwell bulk theory.

\sm

This program has been applied extensively to $2+1$-dimensional gauge theory,
which is realized holographically through $3+1$-dimensional Einstein/Maxwell
theory. Its key solution is the $AdS_4$ black brane with electric charge  $\rho$,
and magnetic field $B$.  This brane solution is known analytically for all $\rho$ and $B$;
its spectrum of small fluctuations may be obtained systematically, and used to
compute physical quantities such as electric and thermal conductivities \cite{Hartnoll:2007ai,Hartnoll:2007ih,Hartnoll:2007ip,Albash:2008eh,Buchbinder:2008dc,Hansen:2008tq,
Caldarelli:2008ze,Hansen:2009xe,Denef:2009yy,Albash:2009wz,Basu:2009qz}.

\sm

While several important condensed matter problems in an external magnetic field,
such as the Quantum Hall Effect and high $T_c$ superconductivity, are driven by
2+1-dimensional physics, it is clearly urgent to obtain results for $3+1$-dimensional
gauge theories as well. For instance, strong magnetic fields are created in collisions at  RHIC, giving rise to observable effects which have been the subject of much recent discussion, e.g.,
\cite{Kharzeev:2007jp,Fukushima:2008xe,Kharzeev:2009pj,Son:2009tf,Yee:2009vw,Rebhan:2009vc}.

\sm

In this paper, we shall present a systematic study of the thermodynamic
properties of $3+1$-dimensional gauge theories with finite electric charge density $\rho$
in the presence of a constant magnetic field $B$.  Their holographic duals should be
electrically and magnetically charged black brane solutions to $4+1$-dimensional
Einstein/Maxwell theory with a Chern-Simons term. The Chern-Simons coupling $k$ captures the strength of the anomaly of the boundary current, and is required to
take a specific value $k=2/\sqrt{3}$ (in our conventions) if the Einstein/Maxwell theory is to
be the bosonic truncation of minimal $D=5$ supergravity; see e.g. \cite{Gauntlett:2003fk}.\footnote{    Whenever $B=0$,
the Chern-Simons coupling is immaterial, and does not enter into the
physical quantities considered here.} It has been proven that any ${\cal N}=1$  superconformal theory with an AdS$_5$ supergravity dual obtained by  compactification from  IIB or
M-theory admits a consistent truncation to  $D=5$ minimal gauged  supergravity \cite{Buchel:2006gb,Gauntlett:2006ai,Gauntlett:2007ma}.  Thus the results we find pertain to a large class of theories, of which ${\cal N}=4$ super Yang-Mills is but one example.

The purely electric solution ($B=0$) is the Reissner-Nordstrom black hole in
$AdS_5$; its
analytic form and thermodynamic properties are well-known.  The existence
of purely magnetic solutions ($\rho=0$) was demonstrated numerically in
a previous paper \cite{D'Hoker:2009mm}, but no analytical solutions are available at present.
Purely magnetic solutions interpolate  between $AdS_5$ and
$AdS_3\times R^2$  near the horizon. As a function
of temperature $T$, their entropy density behaves as $T^3$ for large $T$, and
vanishes as $BT$ for small $T$. These limits agree with $\cN=4$ SYM
calculations at zero gauge coupling, up to factors of $3/4$ and $\sqrt{4/3}$
respectively.  On the  $\cN=4$ SYM side the low temperature thermodynamics is governed
by fermions in the lowest Landau level; an appealing feature of these supergravity
solutions is that they reproduce this low temperature behavior.

Here we wish to extend these results to nonzero $\rho$ and $B$.  Besides their clear usefulness
for applications of AdS/CFT, this investigation has conceptual implications for the status of extremal black holes, as we now pause to discuss.

\subsection{Extremal black holes, Nernst's Theorem, the third law of thermodynamics, and all that}

A striking feature of the Reissner-Nordstrom black brane solution, in any dimension, is
that it possesses a smooth  zero temperature limit with nonzero entropy density.  This extremal solution exhibits a near-horizon AdS$_2 \times R^{D-2}$ region, the existence of which has played a central role in recent holographic descriptions of non-Fermi liquids \cite{Faulkner:2009wj,Liu:2009dm,Cubrovic:2009ye}.  However, a much
discussed cause for concern is that while the extremal entropy apparently plays a crucial
role in this analysis, it is not expected from the point of view of interacting
fermions, nor from the point of view of the dual field theory where the existence of massless charged
bosons suggests that Bose condensation should rule.    One possibility is that the large ground state degeneracy should be understood as an artifact of the large $N$ limit, as discussed in \cite{Denef:2009yy}.  Another is that one should focus on alternative bulk theories where the extremal entropy vanishes, as discussed recently in the case of gauge fields coupled to massless scalars in \cite{Goldstein:2009cv}.
    While this may be the case,
we would like to propose another resolution, based on the results we find for the response
to magnetic fields.

The tension between the existence of extremal black hole entropy and the thermodynamic
behavior of typical systems has been discussed periodically over the years (see e.g. \cite{Wald:1997qp}),
and can be phrased in terms of a clash with Nernst's ``theorem" and the third ``law" of
thermodynamics.  These statements can be expressed in various ways, but essentially they stipulate that
the entropy density $s$  should go to zero at zero temperature (see e.g. \cite{huang}).  Despite the name, this
``law"/``theorem" is actually meant to be a phenomenological observation characterizing the behavior
of observed physical systems.  Indeed, only a moment's thought is required to realize that
it is trivial to concoct theoretical counterexamples based on free systems.  However, such counterexamples are to be thought of as being fine-tuned to an unphysical degree, as any realistic system
will exhibit some degree of interactions, and these will typically lift the ground state
degeneracy.   The relevant question is whether the ground state degeneracy is stable
under adding generic weak interactions or perturbations of the system.  If $\lambda_i$ represent some set of  coupling constants, we should consider
\bea
\lim_{\lambda_i\rightarrow 0} \lim_{T\rightarrow 0} s(\lambda_i,T)~.
\eea
If this limit gives zero for ``typical" couplings $\lambda_i$, then we may conclude that a fine tuning is required to sustain the entropy.

\begin{figure}[h]
\begin{centering}
\includegraphics[scale=0.6]{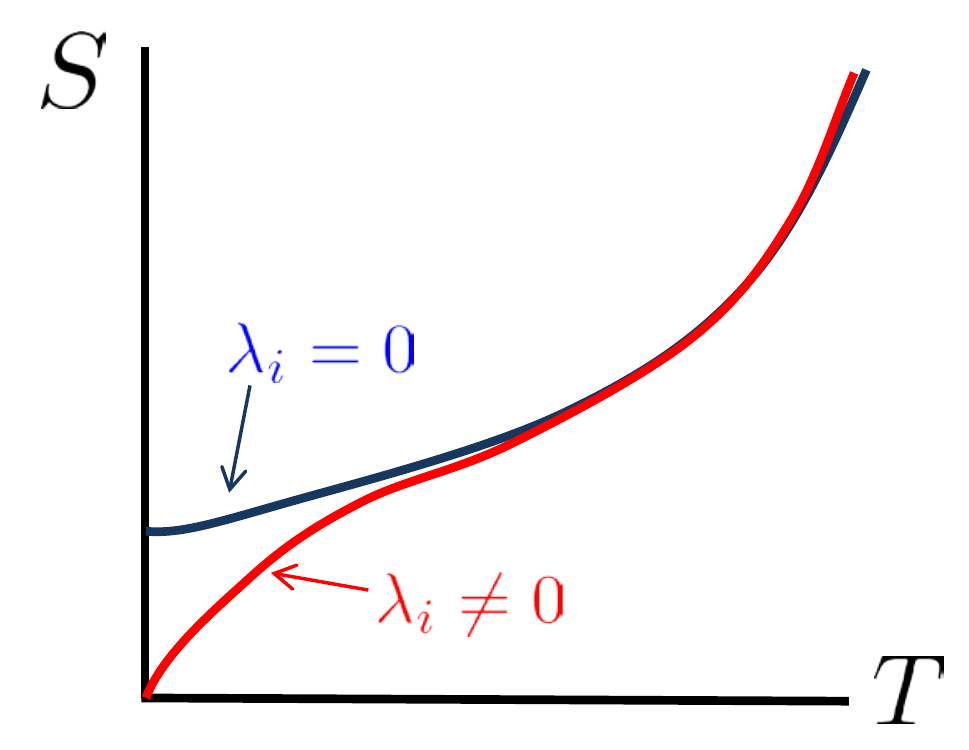}
\caption{Illustration of fine tuning required to maintain nonzero extremal entropy.
Small nonzero couplings $\lambda_i$ lead to no appreciable effect at high temperature, but
cause  the low temperature entropy to flow to zero. }
\label{finetune}
\end{centering}
\end{figure}

Phrased in this way, Nernst's ``theorem" admits a natural formulation in terms of black
holes in the AdS/CFT correspondence.   We should ask whether the extremal black branes
exhibiting finite entropy at extremality are fine-tuned in the same sense when we change
the interactions.  In AdS/CFT we change the Lagrangian of the CFT by changing boundary
conditions for fields in the bulk, and thus we can ask whether the extremal entropy persists
even in the presence of nontrivial boundary conditions for a suitable set of bulk fields.
In a scenario in which the entropy is to be regarded as requiring fine-tuning, we would expect
to see behavior like that in Fig. \ref{finetune}.

In this work we study the case in which $\lambda_i$ corresponds to turning on a constant external magnetic field coupling to the R-current of ${\cal N}=4$ SYM (or other superconformal field theories described holographically), and we will present evidence that the extremal
entropy indeed is unstable  in the above sense under the inclusion of a magnetic field.
Since we proceed numerically, and our numerics break down at very low temperature, we are
not able to follow the entropy all the way down to zero, but the simplest extrapolation suggests
a picture in concordance with Fig. \ref{finetune}.  In further support of this interpretation, we will see that an attempt to construct a finite entropy solution perturbatively in $B$ breaks down  at
very low temperature.    These results suggest that conclusions drawn
from the extremal black brane solution should be viewed with caution, as they can be
drastically affected by even a small (perhaps even arbitrarily small) magnetic field.
It is an interesting question to explore the effect of introducing boundary conditions for
other fields and to study their effect on extremal branes in various dimensions \cite{inprogress}. In this regard, it is worth noting that although the asymptotically AdS$_4$ extremal brane solution
maintains its entropy in the presence of a magnetic field, the question remains regarding more
general perturbations \cite{inprogress}.

\subsection{Summary of results}

\begin{figure}[h]
\begin{centering}
\includegraphics[scale=0.75]{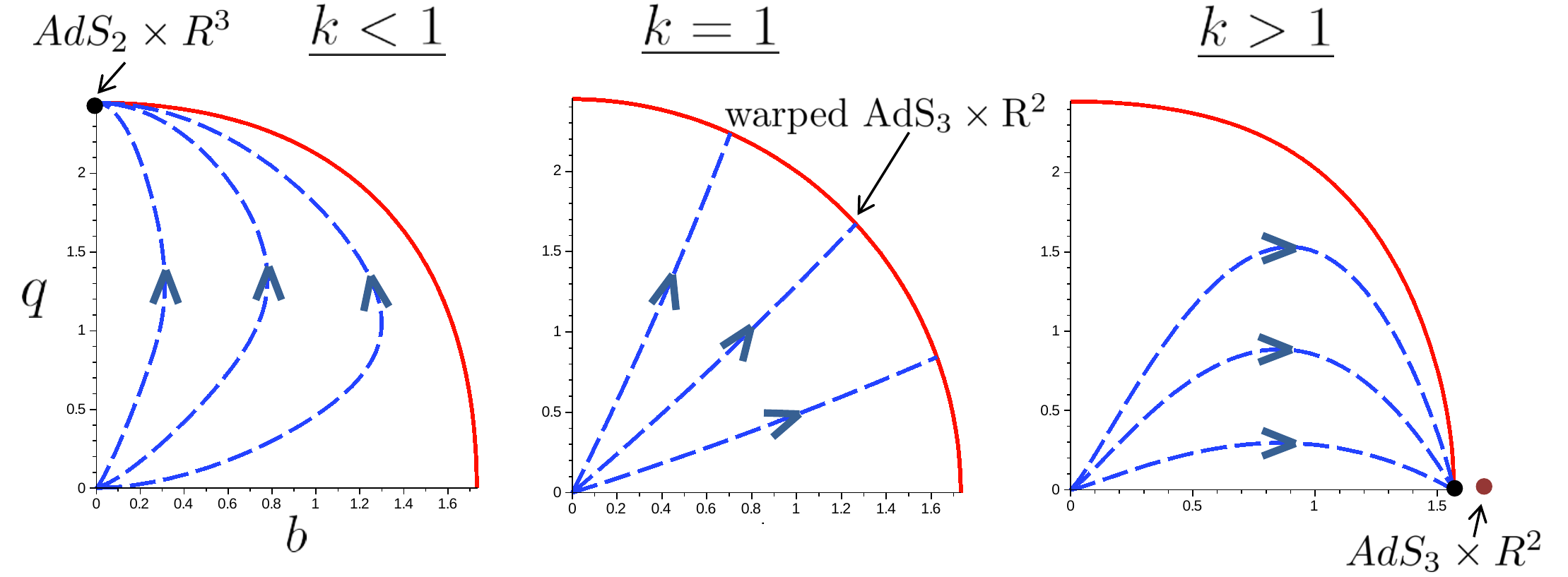}
\caption{Schematic illustration of flows in parameter space for three ranges of
$k$.  Blue lines are flows at various fixed values of $B^3/\rho^2$.  Arrows indicate direction of decreasing temperature.   Red lines indicate the boundary
of allowed $(b,q)$ values where nonsingular solutions are possible.   The near-horizon geometries
at the end points of the flows are indicated in the $k<1$ and $k=1$ cases.  For $k>1$ there
exists an AdS$_3\times R^2$ solution at $b=\sqrt{3}$, indicated by the dot,  but the flows do not reach this point.
In the $k<1$ and $k>1$ cases, near the endpoint of the flow $B^3/\rho^2$ becomes a very sensitive
function of $(b,q)$, depending on the precise direction of approach. }
\label{flow}
\end{centering}
\end{figure}

One of the main results of the present paper is that the low temperature thermodynamics of
solutions carrying nonzero charge density $\rho$ and magnetic field $B$ depends crucially on the
value of the Chern-Simons coupling $k$.  There are three qualitatively distinct cases:  $k<1$, $k=1$,
and $k>1$.  In our conventions, the supersymmetric value is $k =2/\sqrt{3}$, and so falls into the
$k>1$ category.  At high temperatures there is no significant distinction between the three cases.
However, as we take the temperature to zero, holding fixed the dimensionless ratio $B^3/\rho^2$, we find markedly different behavior, as shown in Fig. \ref{flow}.

We parameterize our solutions by $b$ and $q$, which represent the values of the magnetic field
and charge density at the horizon in a particular coordinate system.  They differ from the physical
magnetic field and charge density, which we are calling $B$ and $\rho$; the latter are measured at spatial infinity, and the relation between the two sets of parameters is determined numerically.
For nonsingular solutions, $b$ and $q$ take values in a bounded region, which we can scan over
numerically.  As we lower the temperature holding $B^3/\rho^2$ fixed\footnote{Holding $B$ and $\rho$
fixed independently is not meaningful, as they are dimensionful parameters and are thus changed by a scale transformation.} we flow along lines in the $(b,q)$ plane, in what are essentially  renormalization group flows.  The flows for the various values of $k$ are shown
schematically  in Fig. \ref{flow}. The figures representing our actual numerical data
will be presented in section 7, specifically in Fig. 6.

The flows are driven towards three distinct endpoints, depending on the value of $k$.

\begin{itemize}
\item For $k<1$
we are driven towards $(b=0,q=\sqrt{6})$.  This is the Reissner-Nordstrom black brane with near
horizon geometry AdS$_2 \times R^3$.

\item
For $k=1$ we flow towards the curve $q^2+ 2b^2 =6$; the solutions along this curve have a near-horizon geometry corresponding to warped AdS$_3$ $\times R^2$.
These warped AdS$_3$ geometries have attracted attention recently in the context of topologically massive gravity \cite{Anninos:2008fx} and the Kerr/CFT correspondence \cite{Guica:2008mu}; here we find that they emerge as solutions of $4+1$ dimensional
Einstein-Maxwell theory, and can be connected to asymptotically AdS$_5$ spacetimes.
As we move along the curve the near-horizon geometries interpolate  continuously between AdS$_2\times R^3$  and AdS$_3 \times R^2$.

\item  For
$k>1$, including the supersymmetric value,  we flow towards $(b_c,q=0)$, where $b_c$ is a $k$ dependent number that starts out at
$\sqrt{3}$ for $k=1$ and then decreases with $k$.  To the right of this end  point, at $(\sqrt{3},0)$, is the AdS$_3 \times R^2$ solution that was studied in \cite{D'Hoker:2009mm}.  At the supersymmetric value of the Chern-Simons coupling, $k=2/\sqrt{3}$, the values of $q$ and $b$
are bounded by a critical curve, which to about $0.5 \%$ accuracy, is given by
the relation $q^2 + \a b^2=6$ with $\a \approx  2.44149$  (this value of $\a$ is chosen to give high precision at the endpoint of the curve.)

\end{itemize}

\sm

\begin{figure}[h]
\begin{centering}
\includegraphics[scale=0.75]{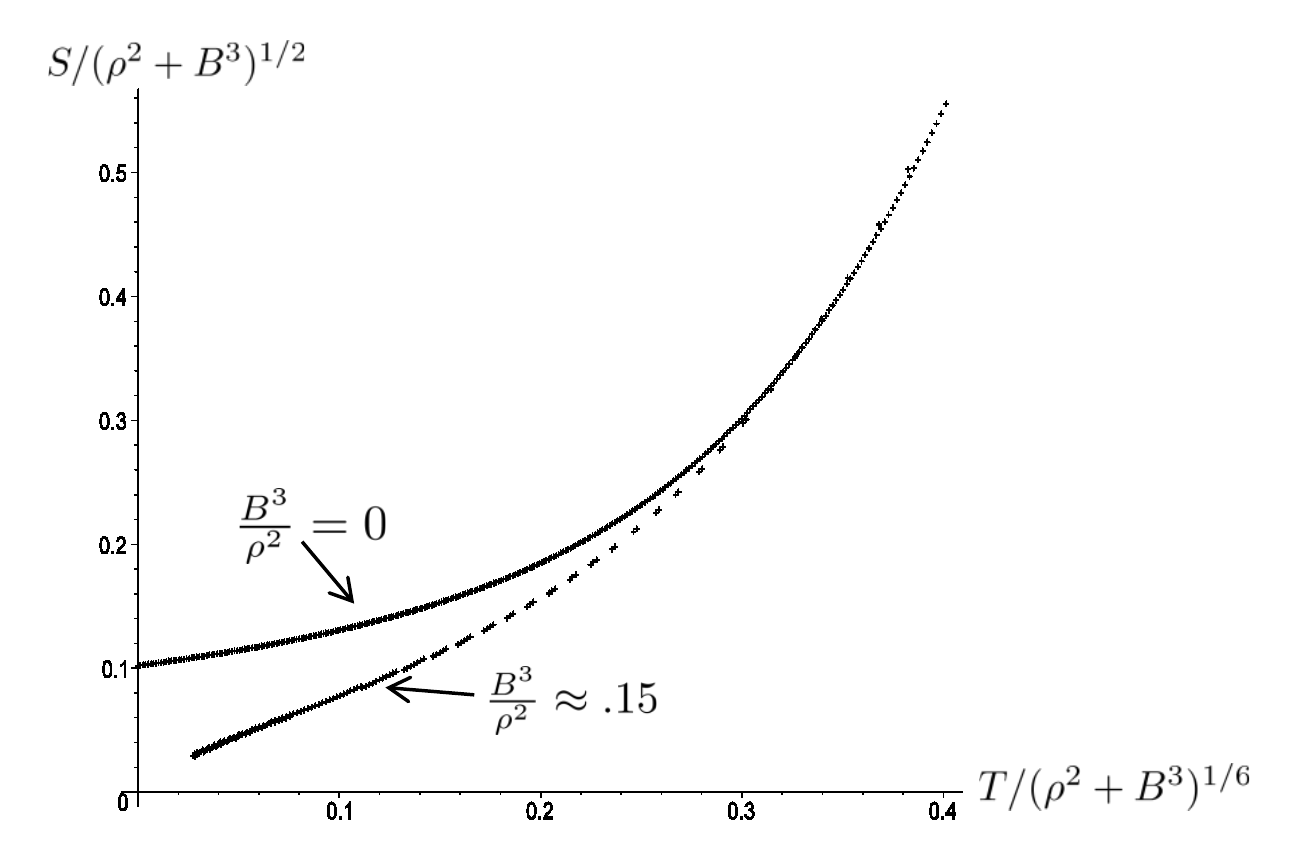}
\caption{Plot of the entropy versus temperature at fixed $B^3/\rho^2=0 $ and $B^3/\rho^2 = .15 \pm .002$, for $k=2/\sqrt{3}$ (supersymmetric value).  The numerical results show that a small $B$ field causes a large drop in the entropy at low temperature.  }
\label{ksusyentropy}
\end{centering}
\end{figure}

As would be expected from the flow diagrams, the behavior of the entropy at low temperature depends on $k$.   In Fig. \ref{ksusyentropy} we show our numerical results for the supersymmetric value
$k = 2/\sqrt{3}$ with $B^3/\rho^2$ fixed at $0$ and approximately $.15$.  We plot dimensionless
versions of the entropy density and temperature, since the dimensionful versions
have no intrinsic meaning.

This plot illustrates
that a small value of $B$ causes a large decrease in the entropy.  This behavior is
representative of the $k>1$ case in general, and  the effect  seems to get more pronounced
with increasing $k$.      Our numerics break down at
low temperatures due to our choice of gauge fixing, and we have stopped our numerics in a regime
where the results are still reliable.  Extrapolating further, it is possible that a
singularity or instability arises at some finite temperature.  However, what cannot happen is that
we end up at a smooth finite entropy extremal black hole, as we will show that no such solution
exists at nonzero $B$ (within our Ansatz, which assumes such properties as translation invariance).

\begin{figure}[h]
\begin{centering}
\includegraphics[scale=0.75]{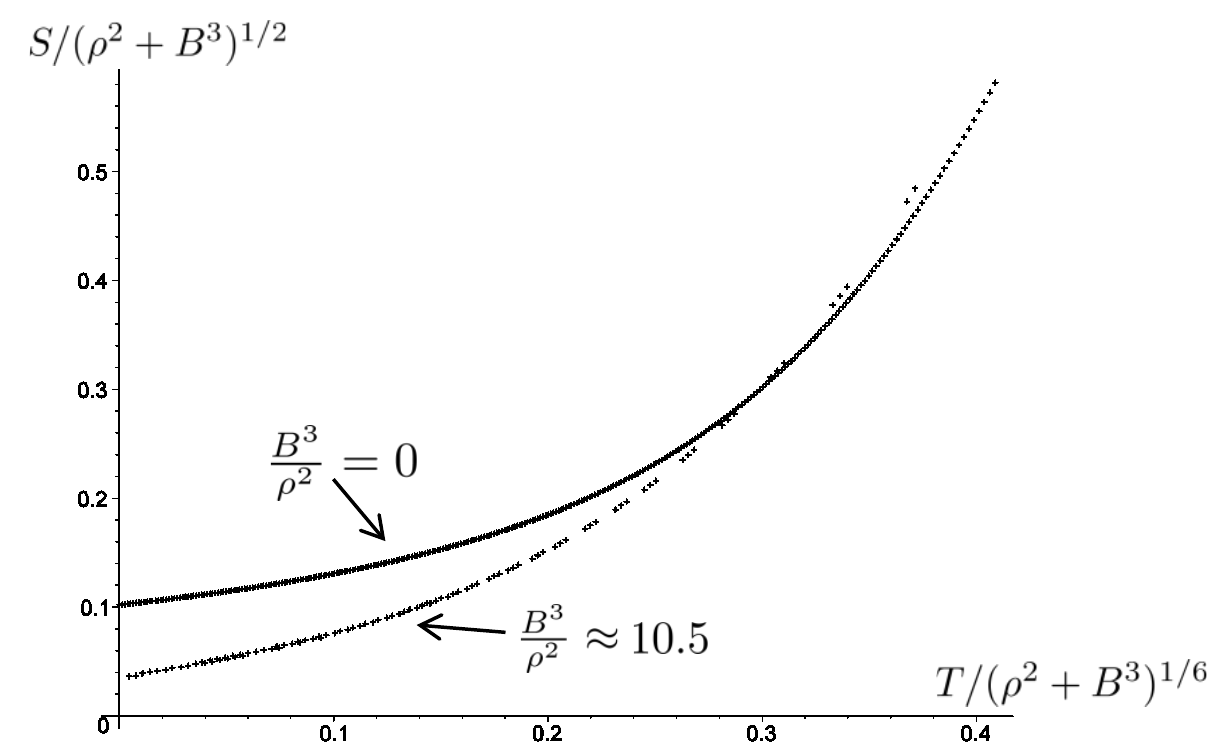}
\caption{Plot of the entropy versus temperature at fixed $B^3/\rho^2=0 $ and $B^3/\rho^2 =10.5 \pm .6$, for $k=0$.  The entropy appears to go to a finite value  (but see the text for comments on the strict zero temperature limit).   }
\label{entropy.k.0}
\end{centering}
\end{figure}

In the $k<1$ and $k=1$ cases the situation is dramatically different, as the entropy appears to go to a finite value (which depends on the value of $B^3/\rho^2$), as shown for $k=0$ in 
Fig. \ref{entropy.k.0}. 
In the $k=1$ case the thermodynamics
is governed by spacelike warped AdS$_3$ black hole solutions, which represent 
exact near-horizon
geometries for our theory.  At strictly zero temperature, for both $k<1$ and $k=1$, the full interpolating solutions acquire a relatively mild singularity at the horizon, unless $B=0$ in which
case we recover the Reissner-Nordstrom solution.   These singularities can be understood from a
perturbative analysis.   Thus, strictly speaking, in these cases there is no smooth
extremal finite entropy geometry, just as there was not in the  $k>1$ case.  However, it may be more physically relevant to
focus on the behavior for small but nonzero temperature, in which case a residual entropy is evident.
\sm

Besides our numerical results, which are valid for arbitrary $B$ and $\rho$, we have carried
out a perturbative analysis of the solutions valid to order $B^2$.   These solutions can be obtained
analytically by methods analogous to those employed in the AdS/fluid dynamics literature, starting with \cite{Bhattacharyya:2008jc}.  Insofar
as they overlap, the perturbative results corroborate our numerical findings.

\sm

The organization of the remainder of this paper is as follows. In section 2,
we briefly review the $D=5$  Einstein/Maxwell theory with Chern-Simons term,
including the definition of the boundary current and stress tensor for asymptotically
AdS$_5$ solutions. In section 3 we
present the Ansatz for uniform electric charge density and constant magnetic field,
and derive the reduced field equations. 
In section 4, regularity and boundary
conditions are discussed both at the horizon and in the asymptotic $AdS_5$
region. Standard AdS/CFT formalism is used to express physical quantities
such as the entropy, chemical potential, currents and stress tensor in terms
of these asymptotic data. In section 5, the near-horizon geometries of our
solutions are constructed analytically, and we discuss the existence of solutions
that interpolate between these and AdS$_5$.
In section 6, a perturbative expansion in powers of $B$ is shown to be
smooth, except for extremal solutions.   In section 7,
numerical results are presented, and a  picture of the phase
diagram is assembled. In section 8 we conclude with a discussion.  
A number of appendices include details of computations omitted in the main text, a 
general discussion on the existence of factorized solutions, and the construction of
solutions with a near-horizon extremal BTZ factor.

\sm

\section{Action, Current, and Stress Tensor}
\setcounter{equation}{0}

The action for 5-dimensional Einstein-Maxwell theory with negative cosmological
constant, and Chern-Simons term, is given by\footnote{Conventions:
$R^\lambda_{~\mu\nu\kappa}=
\p_\kappa \Gamma^\lambda_{\mu\nu}
-\p_\nu \Gamma^\lambda_{\mu\kappa}
+\Gamma^\eta_{\mu\nu}\Gamma^\lambda_{\kappa\eta}
-\Gamma^\eta_{\mu\kappa}\Gamma^\lambda_{\nu\eta}$
and $R_{\mu\nu} = R^\lambda_{~\mu \lambda \nu}$.  }
\bea
\label{action1}
S_{{\rm EM}} = - { 1 \over 16 \pi G_5} \int d^5x \sqrt{-g}
\left ( R + F^{MN} F_{MN} - {12 \over L^2} \right ) + S_{{\rm CS}} + S_{{\rm bndy}}
\eea
where the Chern-Simons action is given by,\footnote{Note that our convention for $k$ differs from that in \cite{D'Hoker:2009mm}:  $k_{\rm{here}}= {3\over 4} k_{\rm{there}}$.  The present
convention has been chosen to simplify the Maxwell equation. By sign-reversal of $A$, 
we are free to choose $k \geq 0$.}
\bea
\label{yb}
S_{{\rm CS}} = {k\over 12\pi G_5} \int  A\wedge F\wedge F
\eea
For the value $k=k_s=2/\sqrt{3}$, the action coincides with the bosonic part of $D=5$
minimal gauged supergravity. In this paper, however,  $k$ will often be kept general,
thus allowing for values different from $k_s$ as well. Boundary terms in the action
are required for the proper renormalization of various physical quantities \cite{Henningson:1998gx,Balasubramanian:1999re,Taylor:2000xw}. In a coordinate
system $(r,x^\mu)$ where $g_{r\mu}=0$ asymptotically for $\mu =0,1,2,3$, the boundary action
$S_{{\rm bndy}}$ is given by,
\bea
\label{yaa}
S_{{\rm bndy}}={1\over 8\pi G_5} \int_{\p M} \! d^4x \sqrt{-\gamma}
\bigg( K -{3\over L}+{L\over 4} R(\gamma)+
{L \over 2}  \left ( \ln {r \over L} \right ) F^{\mu\nu}F_{\mu\nu} \bigg )
\eea
Here, $\gamma_{\mu \nu} $ is the metric induced by $g_{MN}$ on the boundary, and
$K$ is the trace (with respect to $\gamma$) of the extrinsic curvature
of the boundary given by $K_{\mu\nu} =   (\p_r \gamma_{\mu\nu})/(2\sqrt{g_{rr}})$.
Henceforth we set the AdS radius to unity: $L=1$.  The non-diffeomorphism invariant
$\ln r$ term in the boundary  action is needed to remove the divergence associated
with the trace anomaly $T^\mu_\mu \sim  F^{\mu\nu}F_{\mu\nu}$.

\sm

The Bianchi identity is $dF=0$, while the field equations are given by,
\bea
0 & = &  d * F + k F \wedge F
\no \\
R_{MN} & =  &
4  g_{MN} +{1 \over 3} F^{PQ}F_{PQ} g_{MN} -2 F_{MP}F_N{}^P
\eea

\subsection{Boundary Current and Stress Tensor}

For large $r$ the boundary metric of asymptotically AdS$_5$ solutions will behave as
\bea \gamma_{\mu\nu}= r^2 \gamma^{(0)}_{\mu\nu}+ \cdots
\eea
where $\gamma^{(0)}_{\mu \nu}$ is the conformal boundary metric, given here by the flat Minkowski metric.  Similarly, the components of the gauge field $A_\mu$ tangent to the boundary will
go to a constant at large $r$, representing a constant magnetic field pointing in the $x_3$ direction.

By considering on-shell variations of the boundary metric and gauge field, we can define
a boundary current and stress tensor in the familiar fashion:
\bea
\delta S = \int\! d^4x \sqrt{-\gamma^{(0)}}\left( J^\mu \delta A_\mu + {1\over 2} T^{\mu\nu}\delta \gamma^{(0)}_{\mu\nu}\right)
\eea
Specializing to the case of a constant field strength on a flat boundary metric, we have

\bea
\label{yfa}
- 4\pi G_5J^\mu &=&
\Big(r^3 \gamma_{(0)}^{\mu\nu}F_{r\nu}  +{k\over 3}
\epsilon^{\alpha\beta\gamma\mu}A_\alpha F_{\beta\gamma} \Big) \\ \no
8\pi G_5T^{\mu\nu} &=& r^6
\Bigg(-K^{\mu\nu}+K\gamma^{\mu\nu}-3\gamma^{\mu\nu}-2\left(
F^{\mu\alpha}F^{\nu}_{~\alpha}-{1\over 4} F^{\alpha\beta}F_{\alpha \beta}\gamma^{\mu\nu} \right) \ln r               \Bigg)
\eea
where the limit of large $r$ is implied.  For our solutions, the explicit $\ln r$ terms will
cancel logarithmic terms in the metric functions, yielding a finite large $r$ limit for
$T^{\mu\nu}$.

\section{Ansatz and Reduced Field Equations}
\setcounter{equation}{0}

The presence of uniform magnetic field and electric charge density
in the boundary CFT may be achieved by an Ansatz for the bulk fields
which is invariant under translations in $x^\mu$, and space-rotations
around the magnetic field, which we shall take to be pointing in the $x_3$
direction. The Ansatz consistent with these symmetries is given by\footnote{An additional
term of the form $N(r) dx^3 \wedge dt$ must have constant $N$ in view of the
Bianchi identities, and $N=0$ in view of the field equations, and has thus been omitted.}
\bea
\label{anz1}
F = E (r) dr \wedge dt + B dx_1 \wedge dx_2
+ P(r) dx_3 \wedge dr
\eea
for the Maxwell field strength, and by
\bea
\label{anz2}
ds^2 = {dr^2 \over U(r)} - U(r) dt^2 + e^{2V(r)} \left ( dx_1^2 + dx_2^2 \right  )
+ e^{2 W(r)} \left ( dx_3 + C (r) dt \right )^2
\eea
for the metric. The magnetic field $B$ is forced to be constant by the Bianchi identity,
and the functions $E,P,U,V,W$, and $C$ depend only on $r$. Reparametrization
invariance in $r$ has been used to select a coordinate $r$ for which the same function $U(r)$
appears in both $g_{rr}$ and the first factor of $dt^2$.
Rescaling $x_1,x_2$ can be compensated by a constant shift of $V$, while  rescaling
$x_3$ can be compensated by scaling $C$ and shifting $W$.

\sm

In this coordinate system the event horizon is located at $U(r_+)=0$, and the Hawking
temperature is given by
\bea
T= {U'(r_+)\over 4\pi}
\eea

\sm

The above Ansatz is covariant under the following transformation,
\bea
x_3 & \to & x_3 - \a t
\no \\
C(r)  & \to & C(r) + \a
\no \\
E(r)  & \to & E(r) - \a P(r)
\eea
with all other coordinates and fields left unchanged, and for any real parameter $\a$.
We shall refer to this transformation as {\sl $\a$-symmetry}. It may be used, for example,
to set $C(\infty)$ to zero. Note that the combinations $E+C P$ and $C'$ are invariant under
$\a$-symmetry.

The Ansatz  (\ref{anz1}), (\ref{anz2}) is also covariant under boosts in the direction
of the magnetic field.  We can use these boosts to put the solution in the rest frame,
defined by $C(r_+)=0$.  In the dual CFT this corresponds to setting to zero the 
chemical potential conjugate to momentum.  See appendix \ref{AsyBoost} for the details.

\subsection{The reduced Maxwell equations}

The Bianchi identity is automatic on the Ansatz (\ref{anz1}).  The reduced
Maxwell equations are,
\bea
\label{max1a}
\left (  (E + C P) e^{2V+W} \right ) ' + 2 kB P & = & 0
\no \\
\left ( U P e^{2V- W} - C(E +C P) e^{2V+W} \right ) ' + 2 kB E & = & 0
\eea
Both equations may be recast in terms of the $\a$-symmetry invariant
combinations $C'$ and
\bea
\cE \equiv E+PC
\eea
only, by eliminating the derivative of $(E + C P) e^{2V+W} $ in the second equation
using the first equation. One then obtains an alternative form of the reduced Maxwell
equations,
\bea
\label{max2a}
M1 & \hskip .5in & \left (  \cE e^{2V+W} \right ) ' + 2 kB P = 0
\no \\
M2 & \hskip .5in & \left ( U P e^{2V-W}  \right ) '  - C' \cE e^{2 V + W}  + 2 k B \cE = 0
\eea

\subsection{The reduced Einstein equations}

The reduced Einstein equations may be simplified to assume the following final form,
\bea
\label{EQS2}
E1 & \hskip .5in & \left ( C' e^{2V+3W} \right )' = 4 P \cE e^{2 V + W}
\\
E2 & \hskip .5in & U (V'' - W'')  + \Big ( U' + U ( 2 V' + W') \Big ) (V'-W')
\no \\ && \qquad \qquad
= \half (C')^2 e^{2W} - 2 B^2 e^{- 4 V} + 2 U P ^2 e^{-2W}
\no \\
E3 & \hskip .5in & U V'' + U' V' + U V' ( 2 V' + W')
= 4  - { 2 \over 3} \cE^2
-{ 4 \over 3} B^2 e^{-4V} +{2 \over 3} U P^2 e^{-2W}
\no \\
E4 & \hskip .5in & U '' + U' (2 V' + W') -  (C')^2 e^{2W}
=  8  + { 8 \over 3} \cE^2
+ { 4 \over 3} B^2 e^{-4V} +{4 \over 3} U P^2 e^{-2W}
\no
\eea
along with the constraint equation,
\bea
\label{EQS3}
CON & \hskip .5in &
U'(2V'+W') + 2 U (V')^2 + 4 U V' W' + \half (C')^2 e^{2W}
\no \\ && \qquad \qquad
= 12 -2 \cE^2 -2 B^2 e^{-4V} + 2 U P^2 e^{-2W} \hskip 1.5in
\eea
The $r$-derivative of the constraint vanishes by the use of the other
six equations, and may be enforced as an initial condition, as usual.

\section{Asymptotics and initial data}
\setcounter{equation}{0}

The solutions we consider are asymptotically $AdS_5$. Thus, $U(r)$,
$e^{2V(r)}$ and $e^{2W(r)}$ behave as $r^2$ in the limit $ r \to \infty$.
The precise overall normalization depends on the normalization of the
space coordinates $x_1,x_2$ (for $V$) and $x_3$ (for $W$).
There are two natural ways of normalizing this behavior, by parametrizing
either the initial data at the horizon, or the asymptotic behavior as $r \to \infty$.
We shall consider both, and relate their behaviors. The data at the horizon is used for  the numerical analysis, and for computing the entropy and temperature.      The asymptotics
are used to calculate the Maxwell current and the stress tensor.

\subsection{Parametrizing the initial data at the horizon}
\label{parhor}

In the numerical analysis it is important to choose coordinates to remove the gauge freedom.
This can be done by demanding that the solution take a canonical form at the horizon.
By rescaling of $x_1,x_2,x_3$, and  combining an $\a$-transformation and
a boost in the $x_3$-direction, the field strength $F_H$ and the metric $ds_H^2$
at the horizon may be arranged to take the form,
\bea
F_H & = & q \, dr\wedge dt + b \, dx_1 \wedge dx_2
\no \\
ds_H^2 & = & dx_1^2 + dx_2^2 + dx_3^2
\eea
where  $q$ and $b$ are respectively the charge density and the
magnetic field at the horizon (in the coordinates $x_1,x_2,x_3$).
This corresponds to the following initial conditions at the horizon,
\bea\label{inhor}
\cE(r_+) = q \hskip 0.6in U(r_+)=V(r_+)=W(r_+)=C(r_+)=P(r_+)=0
\eea
We will refer to these coordinates as the {\it horizon frame}. 
It remains to specify $V'(r_+), W'(r_+)$, and $C'(r_+)$. These quantities
are generally not independent, but rather follow from the reduced equations
M2, E2, and CON evaluated at the horizon, and we have,
\bea\label{za}
q \left ( C'(r_+) - 2kb \right ) & = &0
\no \\
U'(r_+) V'(r_+) & = & 4 -{2 \over 3} q^2 -{4 \over 3} b^2
\no \\
U'(r_+) W'(r_+) & = & 4 -{2 \over 3} (q^2-b^2) -\half C'(r_+)^2
\eea
The value of $C'(r_+)$ is specified to be $C'(r_+)=2kb$ for $q \not=0$, but
remains an independent free parameter for $q=0$. The quantity $U'(r_+)= 4 \pi T$
is not a genuine initial datum, since the equation CON for $U$ is of first
order. Therefore, genuinely distinct solutions are specified by only
two parameters, for example $T$ and $q$ in units of magnetic field $b$.

If the
temperature $T$ is nonzero, we can always  rescale $t$ to set $U'(r_+)=1$, leaving
the free parameters $b$ and $q$.    Furthermore, we can shift $r$ to set the horizon
at $r_+=1$.

\subsection{Extremal solutions require $bq(k\pm 1)=0$}
\label{noext}

It is now easy to establish a non-existence result that will play an important role in what
follows.  We ask under what conditions can
we have an extremal horizon, $U(r_+)=U'(r_+)=0$.

Assuming that all functions in our Ansatz are well behaved at $r_+$, we can always
work in the horizon frame specified in the previous subsection, in which case the conditions (\ref{za}) apply.   But then it  is easy to see that the assumption of an extremal horizon with nonzero $b$ and $q$ is inconsistent with   (\ref{za})  unless $k^2=1$.
To obtain an extremal horizon we must choose one (or more) of $q=0$, $b=0$, or $k=\pm 1$.
As will be discussed in section \ref{nearhorizon}, these three choices lead to near-horizon  geometries of the form AdS$_2\times R^3$,  AdS$_3\times R^2$, and warped AdS$_3 \times R^2$.   But in the generic case in which none of these conditions is satisfied, finite temperature solutions cannot be smoothly
brought to zero temperature, a feature that we will see explicitly from various points of view.

\subsection{Asymptotic behavior as $r \to \infty$}
\label{physquant}

Starting from the initial data (\ref{inhor}) and integrating our to large $r$, we will find
asymptotically AdS$_5$ solutions with large $r$ behavior
\bea
\label{ub}
U= (r-r_0)^2 + {u_2 \over r^2} + u_2' {\ln r \over r^2} + \cdots  \hskip 0.1in
& \hskip 1in &
C= c_0 + {c_4 \over r^4} + \cdots
\no \\
e^{2  V} = v (r-r_0)^2+ { v_2 \over r^2} + v_2' {\ln r \over r^2} +\cdots \hskip 0.05in
&&
\cE = {e_3  \over r^3}+\cdots
\no \\
e^{2 W} = w (r-r_0)^2+ {w_2 \over r^2} + w_2' {\ln r\over r^2} +\cdots
&&
P= {p_3  \over r^3}+\cdots
\eea
where the dots stand for higher order terms in $1/r$.
Some of the parameters are related to one another by the  field
equations,
\bea
\label{ubb}
w_2 = -  {2 w v_2 \over v} & \hskip 0.7in & u_2' = -{ 2b^2 \over 3v^2}
\no \\
v_2' = {b^2 \over 3v }  \hskip 0.2in && w_2' = -{2b^2 w \over 3v^2}
\eea
In these coordinates the conformal boundary metric is $-dt^2 + v(dx_1^2 + dx_2^2)+w(dx_3+c_0dt)^2$.  As we show in appendix \ref{AsyCoord}, we can always perform a coordinate
transformation to set $v=w=1$ and $c_0=0$, which brings the conformal boundary metric to
the standard Minkowski form.  Further, this can be done while preserving the condition that
we be in the rest frame, defined by $C(r_+)=0$.  The components of the current  in this frame, which we refer to as the {\it asymptotic frame} since it is the relevant one for comparing with the boundary CFT, are computed to be
\bea\label{yfaw}
4\pi G_5 J^t \equiv \rho &=&  \gamma_c(e_3-c_0 p_3) -{2kb \over 3v} A_3|_{\infty} \\ \no
J^{1,2}&=&0 \\ \no
4\pi G_5 J^3 &=& {3\over 4}\gamma_c \left({p_3 \over \sqrt{w}}-\sqrt{w} c_0 e_3\right)
\eea
Here $B$ is the value of the magnetic field in the asymptotic frame, thus identified as the magnetic
field in the CFT, and given by
\bea
B= {b\over v}
\eea
$\gamma_c$ is a Lorentz boost factor (we use the notation $\gamma_c$ to avoid confusion with
the boundary metric $\gamma$), appearing when we transform to the rest frame,
\bea
\gamma_c ={1\over \sqrt {1-w c_0^2} }
\eea

For black hole solutions in which $g_{33}$ remains finite at the horizon, $A_3|_{\infty}$
is arbitrary, and can be adjusted by a constant shift of $A_3$ throughout spacetime.
Its appearance in the expression for $J^t$ is a consequence of the anomaly equation for the
boundary current.\footnote{Given $A_3(t)$ and a constant
magnetic field along $x^3$ we have $\p_t J^t \sim kE \cdot B \sim kB \p_t A_3$.
Integrating gives $J^t \sim kBA_3$, in accord with (\ref{yfa}).}  A similar factor of
$A_t|_{\infty}$ appears implicitly in $J^3$, but since
we have fixed $A_t=0$ at the horizon, the asymptotic value of $A_t$ is determined without
ambiguity.   A nonzero value of $A_3|_{\infty}$ corresponds in the CFT to adding a chemical
potential for $J^3$; it is simplest to set it to zero, and we do that henceforth unless
stated otherwise.

We similarly have expressions for the temperature, entropy density, and chemical potential in
the asymptotic frame:
\bea\label{yfb}
T&=&  {\gamma_c U'(r_+) \over 4\pi}  \\ \no
G_5\left({ S\over {\rm Vol}}\right)\equiv s &=& {1\over 4 \sqrt{v^2 w\gamma_c^2}} \\ \no
\mu &=& {3  \gamma_c v \over 8 kb }
\left(  \sqrt{w} c_0 e_3 - {p_3 \over \sqrt{w}}\right)
\eea

Comparing (\ref{yfaw}) and (\ref{yfb}) we note the simple relation
\bea
4\pi G_5 J^3 = -{1\over 2}kB\mu
\eea
This is the chiral magnetic current \cite{Kharzeev:2007jp,Fukushima:2008xe,Kharzeev:2009pj}:  in the presence of a nonzero anomaly coefficient $k$ and a 
chemical potential, a current is induced parallel to the applied magnetic field. This
result also follows from the anomaly equation, as can be seen by allowing for a slow
variation in $x_3$, differentiating both sides, and identifying $\vec{\nabla} \mu$ with an
electric field.

\subsubsection{Physical quantities}

The global $AdS_5$ solution is invariant under scale transformations,
\bea
x'^\mu =  x^\mu/ \ell ~,\quad\quad r' =  \ell r
\eea
Asymptotically $AdS_5$ solutions inherit this as an asymptotic symmetry,
reflecting the CFT nature of the holographic dual theory. Individual quantities,
such as $B$ and $\rho$, transform under these scalings, just as the coordinates
$x^\mu$ do, and so have no independent meaning.  We should instead look at
scale invariant quantities, which have physical meaning.  This is the same as looking at
dimensionless quantities from the boundary point of view.  Let's write
$O \sim \ell ^p$ to denote the transformation $O' =\ell ^p O$. From the
asymptotic invariance of the field strength $F$ and the metric $ds^2$, we find,
\bea
s \sim \ell^3~,\quad\quad
T\sim \ell~,\quad\quad
B\sim \ell^{2}~,\quad\quad
\rho \sim \ell^{3}~,\quad\quad
\mu \sim \ell
\eea
Any combination behaving as $\ell^0$ is a good physical quantity to compute.

\section{Near-horizon geometries}
\setcounter{equation}{0}
\label{nearhorizon}

For generic assignments of the physical parameters $\rho$ and $B$,
analytical solutions are not available in $AdS_5$  (in contrast with the
$AdS_4$ case where an electric-magnetic duality rotation acting on the $B=0$ solution produces a simple dyonic solution). Even the special case of $\rho=0$ at zero temperature
(and $B\not=0$) does not lend itself to a full  analytical solution \cite{D'Hoker:2009mm}.

\sm

Considerable qualitative and quantitative progress can be made,
however, by solving for the near-horizon geometry of the solutions.
This will be carried out in this section. Especially important will be
the question as to whether, for given $\rho, B$, solutions with
{\sl extremal} near-horizon geometry exist, and whether they can
support an electric field at  the horizon. The existence of these
extremal solutions is key to understanding the low temperature limit.
One important result, which was already established in section \ref{noext} will be that for nonzero $\rho$ and $B$, and for $k\neq 1$ there do not exist smooth, finite entropy, extremal solutions.

\sm

In the low  temperature regime, the full solutions may then then be viewed as  interpolations
between asymptotic $AdS_5$ and these near-horizon geometries.
By numerical study, to be discussed in full in section \ref{numerics}, the regularity
of this interpolation will be verified (except at strictly zero temperature, where singularities develop), and the physical properties of the
solution, such as entropy, temperature, and mass will be evaluated.

A general discussion on the existence of factorized solutions may be found in appendix
\ref{factorized}. 

\subsection{General conditions for the existence of extremal solutions}

At an extremal horizon $r_+$ we have $U(r_+)=U'(r_+)=0$.
Extremal solutions provide a natural boundary of the parameter space
of all solutions.

\sm 

To study their existence systematically, it will be convenient to adopt the horizon frame 
specified in section \ref{parhor}.
We scale the coordinates $x_i$ so that $V(r_+)=W(r_+)=0$,
and  denote the magnetic field in these coordinates by $b$.
Reduced Einstein/Maxwell equations M2, E2, E3, and CON,
in which only $U$ and $U'$ enter on the left hand side, produce
a set of non-trivial constraints,
\bea
\label{extr1}
M2 & \hskip 0.7in & q \left ( C'(r_+) - 2 kb \right )=0
\no \\
E2 && C'(r_+)^2 - 4 b^2 =0
\no \\
E3 && 6 - q^2 - 2b^2=0
\eea
where $q$ is defined to be the electric field at the horizon $q = \cE(r_+)$.
The constraint equation CON is a consequence of E2 and E3,
and thus has been omitted from the above list. Eliminating $C'(r_+)$
using the second equation reduces the system to $qb(k \pm 1)=0$ and
$q^2 + 2 b^2=6$. The solutions are as follows,
\begin{enumerate}
\item If $k \not=\pm 1$, then we have $qb=0$, so that either $q=0$ or $b=0$,
leading to the solutions,
\begin{enumerate}
\item The case $b=0$ and $q=\pm \sqrt{6}$, corresponds to the well-known extremal electrically charged black brane (without magnetic field, and
arbitrary value of $k$). Its near-horizon geometry is $AdS_2 \times R^3$.
\item The case $q=0$ and $b=\pm \sqrt{3}$, corresponds to the extremal
purely magnetic brane (without electric charge, and arbitrary value of $k$), obtained in \cite{D'Hoker:2009mm}.
Its near-horizon geometry is $AdS_3 \times R^2$.
\end{enumerate}
\item If $k = \pm 1$, we shall show below that there is in fact a regular solution
for every assignment satisfying $q^2+2b^2=6$, whose near-horizon geometry
smoothly interpolates between $AdS_2 \times R^3$ (at $b=0$) and
$AdS_3 \times R^2$ (at $q=0$).  These solutions can be generalized to include
finite temperature and momentum, and   correspond precisely to one family of
warped black holes considered in \cite{Anninos:2008fx}.
\end{enumerate}

\subsection{Vanishing magnetic field: $AdS_2 \times R^3$}
\label{RNsol}

We begin by briefly reviewing the well-known black brane solution in $AdS_5$
for $B=0$, charge density $\rho$ and mass $M>0$ (the actual charge and mass densities are
proportional to these), given by the functions
$P=C = 0$, and
\bea
\label{xb}
\cE= {\rho\over r^3} \hskip 0.7in V=W=\ln r \hskip 0.7in
U = r^2 + {\rho^2 \over 3r^4}-{M\over r^2}
\eea
In terms of the radii $r_-\leq r_+$ of the inner and outer horizons, we
obtain a convenient parametrization of $\rho, M$ and $U$,
\bea
\rho ^2 & = & 3 r_+^2 r_-^2 (r_+^2 + r_-^2)
\no \\
M & = & r_+^4 + r_-^4 + r_+^2 r_-^2
\no \\
U & = & { 1 \over r^4} (r^2 -r_+^2)(r^2-r_-^2)(r^2+r_+^2+r_-^2)
\eea
As long as $M^3/\rho^4 \geq 3/4$, the singularity at $r=0$ is protected by a horizon. The near-horizon metric reduces to
\bea
ds_H^2 = {dr^2 \over U_H(r)} - U_H(r) dt^2 + r_+^2 (dx_1^2 + dx_2^2 + dx_3^2)
\eea
with $U_H(r) \sim (r^2-r_+^2)$ for the non-extremal case, and
$U_H(r) \sim  (r^2-r_+^2)^2$ for the extremal case. The near-horizon
geometry is factorized into the space-part which is flat, and an $AdS_2$ factor.
The temperature $T$ and entropy density $s$ of the black brane are given by
\bea
T={(r_+^2-r_-^2)(2r_+^2+r_-^2) \over 2 \pi r_+^3}
\hskip 1in
s = {  r_+^3 \over 4 }
\eea
The black brane becomes extremal as $r_- \to r_+$, so that the temperature
goes to zero, but the charge density and entropy density remain finite, and
related by $s/\rho  = 1/(4\sqrt{6})$.

\subsection{Vanishing charge density: $AdS_3 \times R^2$}

With vanishing charge density, the Maxwell field strength $F$ reduces to the $B$-term
only. The solutions in this case were obtained in \cite{D'Hoker:2009mm}. The Maxwell-Einstein equations
have an analytical solution, given by $\cE=P=C=0$, and
\bea
\label{BTZ1}
U= 3 (r^2 - r_+^2) \hskip 0.7in e^{2V}
={B \over \sqrt{3}} \hskip 0.7in e^{2W} =  r^2
\eea
which represents the product of a (non-rotating) BTZ black hole with
$R^2$.
It was confirmed numerically in \cite{D'Hoker:2009mm} that there exists a family of
regular solutions, parametrized by $T/\sqrt{B}$, which interpolate
between the BTZ black hole of (\ref{BTZ1}) at the horizon, and
$AdS_5$ at $r=\infty$. The entropy of these solutions tends to zero as
$T \to 0$, while the physical  magnetic field $B$ is kept fixed.

\sm

More generally, we can have nonextremal rotating BTZ black holes, whose metric is given by
$e^{2V}=B/\sqrt{3}$, and
\bea
\label{BTZ2}
U= 3 {(r^2-r_+^2)(r^2-r_-^2) \over r^2} \hskip 0.7in e^{2W}=r^2 \hskip 0.7in
C= -\sqrt{3} {r_+ r_- \over r^2}
\eea
with $\cE=P=0$.
A useful alternative parametrization of the rotating BTZ solution  is given by,
\bea
\label{BTZ3}
U= 12 (r-r_+)(r-r_-) \hskip 1in C = 2 \sqrt{3} (r-r_+)
\eea
and  $V=W=\cE=P=0$. Note that both forms admit a smooth extremal limit.

\subsection{$k=1$: warped AdS$_3$ black holes}
\label{warped}

As shown above, there is a special value of the Chern-Simons coupling,
namely $k = \pm 1$ (recall that this is less than the value required for supersymmetry), for which there exist extremal solutions for any
$q,b$ satisfying the relation,
\bea
q^2 + 2 b^2 =6
\eea
Furthermore, there is a simple nonextremal generalization.   The solution is given
by $V=W=P=0$, and
\bea
\cE &=& q   \\ \no  U&=&  12 (r-r_+)(r-r_-)
\no \\
C &=& 2b (r-r_+)
\eea
We have used $\a$-symmetry to set $C(r_+)=0$. The metric
and field strength are then  given by,
\bea
\label{inter1}
ds^2 & =& {dr^2 \over U(r)}-U(r) dt^2 + \Big(dx_3 +2b(r-r_+)dt\Big)^2 + dx_1^2 +dx_2^2
\\ \no
F&=& q \, dr \wedge dt + b \, dx_1 \wedge dx_2
\eea
The extremal limit is given by taking $r_- = r_+$ as usual.

These solutions can be identified with the ``self-dual" solutions described in section 6.1.1 of
\cite{Anninos:2008fx}, where we make the identification
\bea
\nu^2 = {3b^2 \over 12-b^2}
\eea
Note that as $b^2$ ranges over its allowed values between $0$ and $3$, $\nu^2$ ranges between
$0$ and $1$.   The equivalence can be seen most easily by comparing our metric with
eqtn. 1.2 of \cite{Compere:2009zj}.  Under the identifications
\bea
\Phi = {12l^2 \over 3+\nu^2}t~,\quad\quad T = \sqrt{ {12l^2 \over 3+\nu^2}}x_3~,\quad\quad R=r
\eea
the metrics are seen to be proportional.   Unlike in \cite{Compere:2009zj}, we do not
compactify $\Phi \sim t$.  Our metric has no closed timelike curves or other pathologies.

These solutions also arise in a context  closely related to ours, namely
$2+1$ dimensional Einstein-Maxwell-Chern-Simons theory (see \cite{Banados:2005da} for solutions which are the analytic continuation of these).  This can be understood from the fact that if
we reduce our theory down to three dimensions along $x_{1,2}$, then we recover the equations
of $2+1$ dimensional Einstein-Maxwell-Chern-Simons theory coupled to a massless scalar field.
The condition for the scalar field to take a constant value, representing a constant value of $V$, is precisely the condition $q^2 +2b^2 =6$ that we found above.
It is curious that these solutions exist only at the special value $k=\pm 1$.

\subsection{Existence of interpolating solutions}
\label{interp}

As we have seen, for nonzero values of $\rho$ and $B$, and $k\neq \pm 1$, there do not
exist smooth zero temperature solutions under the assumptions of our Ansatz (it is possible
that such solutions do exist  if one, for example, relaxes the condition of translation invariance).  So in these cases, if we start from an asymptotically AdS$_5$ solution at finite temperature, as we lower the temperature some of the functions in our solution will start to
diverge;  we will see this as a breakdown of our numerics.

This leaves the question of what happens in the zero temperature limit in the cases for
which there do exist candidate extremal horizons.   In the case of $B=0$ and nonzero
$\rho$ (the value of $k$ is immaterial in this case), the answer is that we end up at the usual
AdS$_5$ extremal Reissner-Nordstrom solution.  This will turn out to be the only case in which
we find a truly nonsingular extremal solution.

For $\rho=0$ and nonzero $B$ ($k$ again drops out of the discussion if one notes that for $q=0$ the equation M2 appearing in (\ref{extr1}) becomes trivial ) interpolating solutions were constructed numerically in \cite{D'Hoker:2009mm}.
At low temperatures the near-horizon geometry approaches AdS$_3\times R^2$, but as will be
discussed momentarily a singularity develops in the full interpolating solution at strictly
zero temperature.

For nonzero $\rho$ and $B$, but $k=\pm 1$, we have candidate near-horizon geometries corresponding to warped AdS$_3$ black holes times $R^2$, and these have a smooth extremal limit. At any nonzero temperature, our numerics will establish the existence of solutions
smoothly interpolating between these near-horizon geometries and AdS$_5$.  As the temperature
is taken to zero the entropy remains finite, but nevertheless a singularity develops at the
horizon, for reasons that can be seen as follows.

To construct a candidate extremal interpolating solution, we can start with the exact near
horizon extremal warped black hole geometry, and then introduce a perturbation that grows
near the boundary, representing the change in asymptotic boundary conditions taking us
towards AdS$_5$.   This perturbation is obtained by solving the equations obtained by
linearizing around the near-horizon solution.

For our near-horizon solution we have
\bea
U&=& 12 (r-1)^2 \\ \no
C&=&  2b(r-1) \\ \no
\E &=&q \\ \no
V&=& W=P=0 \\ \no
\eea
with $q^2+2b^2=6$.  Note that we have used the freedom to rescale $r$ to set the horizon
at $r_+=1$.

Now we linearize around this solution, denoting the perturbations by lower case letters
($\eps$ denotes the perturbation of $\E$).
Plugging in we find the following equations to linear order
\bea
M1:~~&& \delta \eps' +q(2v'+w')+2b p =0 \\ \no
M2:~~&& \left(12(r-1)^2 p\right)'-q c'-2qb(2v+w)=0 \\ \no
E1:~~ && c'' +2b(3w'+2v')-4qp=0 \\ \no
E2:~~ &&\left( 12(r-1)^2 (v'-w')\right)'-8b^2 v -2b c' -4b^2 w=0 \\ \no
E3:~~&& \left( 12(r-1)^2 v'\right)'  -{16 \over 3}b^2 v +{4\over 3}q \eps =0 \\ \no
E4:~~&& 3u'' +72(r-1)(2v'-w')+8b^2(2v-3w)-12bc'-16 q \eps =0 \\ \no
CON:~~&& 24(r-1)(2v'+w')+4b^2(w-2v)+2bc'+4q \eps =0
\eea
We impose the boundary conditions
\bea
u(1)=u'(1)=v(1)=w(1)=c(1)=c'(1)=p(1)=\eps(1)=0
\eea
It is fairly straightforward to solve these equations iteratively.  Of most
relevance are the resulting expressions for $v$, $\eps$ and $p$, which are
\bea
v &=& a_1 (r-1)^\alpha \\ \no
\eps &=& 3q a_1 (r-1)^\alpha \\ \no
p &=& -{3q \alpha a_1 \over 2b}(r-1)^{\alpha-1} -{q\over 2b}a_2
\eea
where $a_{1,2}$ are integration constants, and
\bea
\alpha = -{1\over 2}+ {\sqrt{81-8b^2}\over 6}
\eea
If $q^2\neq 6$ then $b^2 >0$, in which case $\alpha<1$, and then we see that $p$
diverges at the horizon.   Thus to have a smooth solution we are forced to set $a_1=0$.  But this means that $v=\eps =0$, and from here it follows rapidly that the entire solution is
just the original near-horizon geometry we started from.

That the divergence in $p$ represents a physical singularity can be seen by considering
an infalling observer.  One finds that such an observer sees a diverging physical field
strength at the horizon.

We therefore do not expect to find a smooth extremal interpolating solution when $q\neq 0$,
even for $k=\pm 1$ when smooth near-horizon geometries do exist.
In our numerics, as we lower the temperature we indeed find that $P$ begins to diverge
at the horizon in precisely the manner described above.   Nevertheless, since the metric
components have well defined limits (though not their derivatives) we find that the
entropy appears to smoothly approach a finite value.

This leaves the case of nonzero $B$ but $\rho=0$, which was studied in \cite{D'Hoker:2009mm}.
There solutions were found numerically that interpolated between near-horizon 
BTZ $\times R^2$ and AdS$_5$.    At  any finite temperature these interpolating solutions are smooth, but
it follows from the above that a mild singularity develops at strictly zero temperature.
For these solutions $k$ does not appear in the field equations, and so we can freely
set $k=\pm 1$, in which case we can compare with the linearized analysis just described by
setting $q=0$.  For $q=0$ we have that $p=0$, and so we avoid the divergence in that
quantity.  However, we have that $v\sim (r-1)^\alpha$, $\alpha = -1/2+\sqrt{57}/6$.
Since $\alpha<1$, the first derivative of $V$ will diverge at  the horizon, presumably
indicating a singularity. 

Actually, one special case remains.  When $\rho=0$, rather than considering the
zero temperature limit of BTZ corresponding to pure AdS$_3$, we can look for solutions
with near-horizon geometry given by a finite entropy extremal BTZ solution.  These
can be constructed by a fairly standard  construction, as we describe in
appendix D.   The resulting interpolating solutions exhibit the same singularity
at extremality as above; namely, derivatives of $V$ diverge at the horizon.

\section{Perturbation theory in powers of $B$}
\setcounter{equation}{0}

In this section, we shall construct solutions perturbatively in powers of $B$
around the analytically known solution for $B=0$ and arbitrary charge density
$\rho$ and mass $M$. We expect this expansion to be reliable for $T^2 \gg B$.
For small $T$, however, the $B=0$ solution is close to extremal,
and we know from the conditions of extremality of (\ref{extr1})
that no extremal solutions exist with $B \not=0$ and $\rho \not= 0$,
unless we also have $k=\pm 1$. Thus, for $k \not=\pm1 $, we expect
perturbation theory in $B$ to break down near $B \sim T^2$. For $k = \pm 1$ 
the behavior is better, but we find that a mild singularity still results.  
The structure of the computation is very similar to the long wavelength
fluid dynamics from gravity analysis in \cite{Bhattacharyya:2008jc}.
This is because a weak magnetic field corresponds to a slowly varying gauge
field.

\sm

The functions $E(r), U(r), V(r), W(r)$ are
even in $B$, while the functions $P(r)$ and $C(r)$ are odd in $B$.
Here, we shall expand up to order $B^2$ included, so that,
\bea
\label{xa}
U=U_0 +B^2U_2 ~~ & \hskip 1in & E=E_0 +B^2E_2
\\ \no
V=V_0 +B^2V_2 ~~ && C=BC_1
\\ \no
W=W_0 +B^2W_2 && P = BP_1
\eea
The zeroth order solution coincides with (\ref{xb}), and is given by
$P_0=C_0=0$, and
\bea
\label{xb1}
E_0= {\rho\over r^3}
\hskip 0.7in
V_0 = W_0 = \ln r
\hskip 0.7in
U_0 = r^2 + {\rho^2 \over 3r^4}-{M\over r^2}
\eea
The horizons of $U_0$ will be denoted by $r_\pm$.
The boundary conditions include $C_1(r_+)=C_1(\infty)=0$,
together with the requirement that $V_2$ and $W_2$ fall off faster than $r^{-2}$
as $ r \to \infty$.
To separate out the spin zero (scalar) and spin two (tensor) perturbative
corrections, it will be convenient to introduce the combinations,
\bea
\label{xk}
S_2=2V_2+W_2 \hskip 1in  T_2 = V_2-W_2
\eea
Equations M2 and E1 are odd in $B$, and thus have contributions only
to first order in $B$, while all other equations, M1, E2, E3, E4, and CON
are even in $B$ and thus have only second order contributions.
We begin by solving M2 and E1 first.

\subsection{Spin one sector}

To order $B$, equation M2 is given by,
\bea
\label{xd}
\left ( U_0 r P_1-\rho C_1 \right )'=-2 {k\rho \over r^3}
\eea
It may be readily integrated, and the integration constant is fixed by the
boundary conditions at the horizon, $U_0(r_+)= C_1(r_+)=0$.
As a result, $P_1$ is given in terms of $C_1$ by,
\bea
\label{p1}
P_1 (r) = {\rho \over r U_0(r)} \left ( C_1(r) + {k \over r^2} - { k \over r_+^2} \right )
\eea
Note that the function $P_1$ is automatically smooth at the horizon.
To obtain $C_1$, we substitute the solution $P_1$ into equation E1.
Using the special form of the function $U_0$, this equation may be
recast as follows,
\bea
\label{xf}
\left(rU_0^2 \left({r^2 C_1\over U_0}\right)'\right)'
={4k\rho^2 \over  r^5}-{4k \rho^2  \over r^3r_+^2}
\eea
In this form, the equation may be solved by two successive integrations,
producing two integration constants. These constants are fixed uniquely
by the requirements that $C_1(r_+)=C_1(\infty)=0$, and we find,
\bea\label{xh}
C_1 (r)= - k \rho^2 {U_0(r)\over r^2 } \int_{\infty}^r {dr' \over r'U_0^2(r')} \left(
{1 \over r'^2}-{1 \over r_+^2} \right )^2
\eea
Thus, the  functions $P_1$ and $C_1$ are uniquely determined
by the boundary conditions.

\subsection{Spin two sector}

Equation E2 to order $B^2$ gives,
\bea
\label{xk1}
\left(r^3 U_0 T_2'\right)'= {1\over 2} r^5 (C_1')^2+2rU_0(P_1)^2 -{2\over r}
\eea
The terms on the right hand side are known from the solution in the spin
one sector. This equation may be solved by two successive integrations.
The two resulting integration constants may be fixed by demanding
smoothness of $T_2'$ at the horizon,  and the vanishing of $T_2$
at infinity, so that $T_2$ is uniquely given by,
\bea
\label{xm}
T_2(r) = \int_{\infty}^{r}dr'' {1\over r''^3 U_0(r'')}\int_{r_+}^{r''}
dr' \left(  {1\over 2} r'^5 (C_1')^2+2r'U_0(P_1)^2 -{2\over r'}\right)
\eea

\subsection{Spin zero sector}

The functions $E_2$, $U_2$ and $S_2$ all correspond to scalar perturbations.
The linear combination $3\times E3 - CON$ gives an equation for $S_2$ in terms
of $C_1$ and $T_2$,
\bea
U_0 S_2'' +{2 \over r} U_0 S_2'
+ U_0 T_2'' + { 3 \over r} U_0 T_2' + 3 U_0' T_2' = \half r^2 (C_1')^2 - { 2 \over r^4}
\eea
Eliminating $T_2$ using (\ref{xm}), the remaining equation can be recast in the
form, $(r^2 S_2')'=-2P_1^2$, which may be solved by two successive integrations.
We need $S_2$ to fall off faster than $1/r^2$ at infinity to preserve the
boundary. The solution is therefore uniquely fixed to be,
\bea
\label{xp}
S_2(r) = 2  \int ^r _\infty dr' \left ( { 1 \over r} - { 1 \over r'} \right ) P_1(r')^2
\eea
Next, equation M1 determines $E_2$,
\bea\label{xr}
(r^3 E_2)' + (\rho S_2)'+(r^3 P_1 C_1)' +2kP_1=0
\eea
The integration constant can be reabsorbed into $\rho$, and so we have
\bea\label{xs}
E_2 (r) = -{\rho \over r^3}S_2 -P_1 C_1 -{2k\over r^3} \int_{\infty}^r dr'
~P_1(r')
\eea
Equation E4 determines $U_2$. It may be expressed as $(r^3 U_2')' = X$,
with $X$ given by,
\bea
\label{xu}
X (r)= - r^3 U_0' S_2' +{16\rho  \over 3}(E_2 +C_1 P_1)
+r^5 (C_1')^2+{4 r U_0 P_1^2 \over 3}+{4\over 3r}
\eea
The solution which goes to zero at infinity is
\bea
\label{xv}
U_2(r) = \int_\infty^r {dr'' \over r''^3}\int_{r_+}^{r''} dr' X(r')
-{a_3\over 2r^2}
\eea
where  $a_3$ is an integration constant.
Finally, the constraint equation may be checked to hold at $r\to \infty$
to leading order. This will guarantee that it is obeyed throughout.

\subsection{Asymptotic behavior of the perturbative solution}

The full perturbative solution is now fixed.  The free parameters are:  $B$, $\rho$,
$M$, where we're not counting $a_3$ since it can be absorbed into $M$, nor $r_+$
since it is a function of $M$ and $\rho$.  As a result, the asymptotic behavior of these
functions can now be computed, and we find,
\bea
\label{xj}
C_1(r)  & = & { k\rho^2 \over 4 r_+^2 r^4}   \hskip 1.4in c_4 = {k \rho^2 B \over 4 r_+^2}
\no \\
P_1 (r) & = &   -{k\rho \over r_+^2 r^3 }  \hskip 1.4in p_3 = - {k \rho B \over r_+^2}
\no \\
T_2 (r) & = & {\ln r \over 2 r^4}  \hskip 1.6in v=1
\no \\
S_2 (r) & = & -{k^2 \rho^2 \over 15 r_+^4 r^6} \hskip 1.3in w=1
\no \\
E_2 (r)& = &  -{k^2 \rho \over r_+^2 r^5} \hskip 1.45in e_3=\rho
\no \\
U_2 (r)& = & -{2\ln r \over 3 r^2}-{1\over 3 r^2}-{a_3\over 2r^2}
\hskip 0.5in u_2= - M + \left ( { 2 \over 3} \ln r_+ -{1 \over 3} - {a_3 \over 2} \right )
\eea

\subsection{Regularity of the perturbative expansion}

For the perturbative expansion around the non-extremal black brane with $r_- < r_+$,
the functions $C_1$ and $P_1$ fall off fast as $r \to \infty$, and are smooth at the
outer horizon $r_+$, as well as at all other values of $r > r_+$. As a result,
the integrals giving $S_2, T_2, E_2 $ and $U_2$ are rapidly convergent,
and define regular functions throughout.

\sm

The perturbative expansion around the extremal black brane with $r_-=r_+$, however,
is not, generally, well-behaved. For $r_-=r_+$, the function $C_1(r)$ is given analytically by
\bea
C_1 (r) = - {k \over 3 r_+^2}  {U_0(r)\over r^2 }
\left [
{2r_+^2  \over r^2 + 2r_+^2} - { r_+^2 \over r^2 - r_+^2}
 + { 1 \over 3} \ln \left ( { r^2 - r_+^2 \over r^2 + 2r_+^2} \right )\right ]
\eea
The functions $C_1$ and $C'_1$ are smooth throughout. The function $P_1$
is given analytically by
\bea
P_1(r) = - { k \rho \over r_+^2 r^3}
\left [ {2r_+^2  \over r^2 + 2r_+^2} + { 3r^2 + 2 r_+^2 \over 3(r^2 + 2 r_+^2)}
 + { 1 \over 3} \ln \left ( { r^2 - r_+^2 \over r^2 + 2r_+^2} \right )\right ]
\eea
and exhibits a logarithmic singularity as $ r \to r_+$. This singularity is integrable
in the formulas giving the perturbation functions $S_2, T_2, U_2$ and $E_2$.
As a result, the functions $S_2, U_2, E_2$ are smooth, while the function
$T_2$ has a logarithmic singularity as $ r \to r_+$, given by
\bea
T_2(r) \sim  \left({ k^2 -1 \over 6 r_+^4}\right) \ln (r-r_+)
\eea
To summarize, in the extremal limit, the electric current density $P$, and the
tensor perturbation of the metric $T_2$ both diverge logarithmically at the
horizon, and the solution is not globally smooth.

\subsection{Perturbative calculation of entropy, temperature, and mass}

Having normalized the metric at $r=\infty$ to be the conformally standard Minkowski metric,
(\ref{bnd1}) with $v=w=1$, the same metric at the horizon then reads,
\bea
ds_H^2 = r_+^2 e^{2 V(r_+)} (d\ti x_1^2 + d\ti x_2^2) + r_+^2 e^{2W(r_+)} d\ti x_3^2
\eea
The perturbative corrections to the entropy density, temperature, and mass
to order $B^2$ may then be deduced from the metric functions as follows.
The entropy density $s$ is given by
\bea
s = s_0 \left ( 1 + B^2 S_2(r_+) \right )
\eea
Similarly, the temperature $T$ and mass $M$ are given by
\bea
T & = & T_0 + { 1\over 4 \pi}  B^2 U_2'(r_+)
\no \\
M & = & M_0 + B^2 \left ( {1 \over 3} -{2 \over 3} \ln r_+
+ \half \int _{r_+} ^\infty dr' \left ( X(r') - { 4 \over 3 r'} \right ) \right )
\eea
Here, $s_0$, $T_0$, and $M_0$ are respectively the entropy density,
temperature and mass of the $B=0$ black brane. Note that it follows from the form of
(\ref{xp}), that $S_2(r_+)<0$, so that the correction to the entropy density
is always negative.

\subsubsection{Eliminating $r_+$-dependence}

The position of the outer horizon, $r_+$, has dimension and hence no direct
physical meaning. It may be thought of as setting the overall scale, and
plays the role of $\ell$ above. Therefore, any physical quantity must be
independent of $r_+$. Thus, we shall introduce dimensionless coordinates
and quantities, such as
\bea
x \equiv r/ r_+ \hskip 1in \lambda \equiv r_- / r_+
\eea
The $r_+$-dependence may now be isolated in each one of the functions
that enter perturbation theory. For $C_1$ and $P_1$ we define the dimensionless
functions $\hat C_1(x)$ and $\hat P_1(x)$ by,
\bea
C_1(r) = k \rho ^2  \hat C_1(x) / r_+^8
\hskip 1in
P_1(r) = k \rho  \hat P_1(x)/ r_+^5
\eea
We also define the dimensionless functions $\sigma (\lambda)$ and $\tau(\lambda)$ by,
\bea
S_2 (r_+) \equiv  \sigma (\l ) / r_+^4
\hskip 1in
U_2'(r_+) \equiv  \tau (\l ) / r_+^3
\eea
and we use $\rho$ as a physical quantity that sets the scale for $r_+$,
\bea
\rho ^2 = r_+^6 \nu(\l )^6 \hskip 1in \nu (\l )^6 = 3 \l ^2 (1+\l ^2)
\eea
so that $r_+ = \rho ^{1/3} /\nu(\l )$. The expressions for entropy density and temperature
normalized to the physical dimensionful quantity $\rho$ are as follows,
\bea
{s \over \rho}  & = & { 4 \pi \over 3 \nu(\l )^3}
 \left ( 1 +  { B^2 \over \rho^{4/3}} \sigma (\l ) \nu(\l )^4 \right )
\no \\
{ T \over \rho^{1/3}} & = & { 1 \over 2 \pi \nu(\l )}
\left ( (1-\l ^2)(2+\l ^2) +  { B^2 \over 2 \rho^{4/3}}   \tau(\l ) \nu(\l )^4 \right )
\eea

\subsubsection{Calculating the dimensionless functions}

The dimensionless functions $\hat C_1(x)$ and $\hat P_1(x)$ may be readily
computed (analytically) from (\ref{xh}) and (\ref{p1}), and used to evaluate
the dimensionless functions $\sigma(\l)$ and $\tau(\l)$, given by
\bea
\sigma (\l ) & = & - 2 k^2 \nu(\l)^6 \int _1 ^\infty dx \left (1 - { 1 \over x} \right )\hat P_1(x)^2
\no \\
\tau (\l) & = &  -{4 \over 3} + k^2 \l ^2 (1+\l ^2) \hat \tau(\l )
\eea
where $\hat \tau(\l)$ is given by
\bea
\hat \tau(\l )
& = &  -{9 \over 2}  \l ^2(1+\l ^2) \hat C_1'(1)^2
- 9  \l ^2 (1+\l ^2) \int _1 ^\infty dx \, x^3 \hat C_1'(x)^2
-24 \int _1 ^\infty dx \hat P_1(x)
\no \\ &&
+ 8  \int _1 ^\infty {dx \over x^4}\hat P_1(x)
+  \int _1 ^\infty dx \hat P_1(x)^2 \left ( 8  x - 4 \l ^2 (1+\l ^2) {1\over x^5} \right )
\\ &&
+ 2(1-\l ^2)(2+\l ^2) \int _1 ^\infty dx \hat P_1(x)^2
-72 \l ^2(1+\l ^2) \int _1 ^\infty dx \left ( 1-{1\over x} \right ) \hat P_1(x)^2
\no
\eea
The functions $\sigma (\l)$ and $\hat \tau (\l)$ may be evaluated numerically,
using the analytic expression for $\hat C_1(x)$ and $\hat P_1(x)$. The results are
as follows.

\sm

Numerical evaluation of the integral entering $\sigma(\l)$ shows very little dependence
on $\lambda$ throughout  the interval $\lambda \in [0,1]$, and may be well approximated
there by the average value of $0.015$ (specifically, its value drops uniformly from
0.01505 at $\l=0$ to 0.01490 at $\l=1$). As a result, we have the approximate formula,
\bea
\sigma (\l)  \sim -0.090 \times k^2 \l ^2(1+\l ^2)
\eea
for $\lambda$ throughout the interval $[0,1]$.

\sm

Numerical evaluation of the function $\hat \tau(\l )$ produces a dependence  given
in figure \ref{TauHat} below. We record the end point values,
\bea
\hat \tau (0) = 8.605 & \hskip 1in & \tau (0) = - 1.333
\no \\
\hat \tau (1) = 4.527 & \hskip 1in & \tau(1) = -1.333 + 9.054 \times k^2
\eea
and for the supersymmetric value $k^2=4/3$, we have $\tau(1)= 10.739$.

\begin{figure}[h]
\begin{centering}
\includegraphics[scale=0.35]{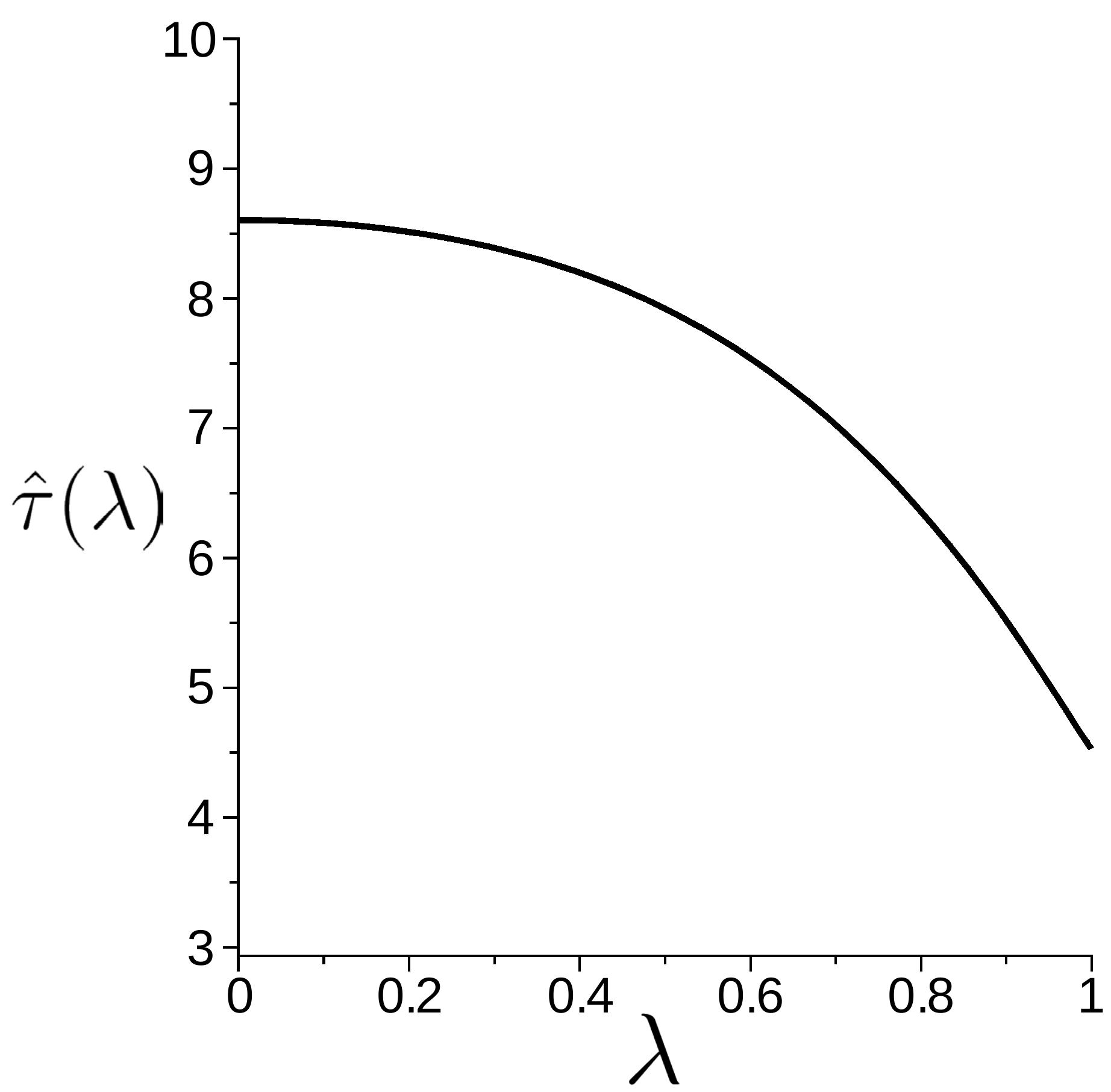}
\caption{ The function $\hat \tau (\l )$.}
\label{TauHat}
\end{centering}
\end{figure}

\subsection{Physical interpretation of the perturbative corrections}

\begin{itemize}
\item If $\tau (1) >0$, then $T$ is non-zero and positive at the extremal value $\l =1$.
We can extract an estimate for the minimum temperature under the assumption that
$B^3/\rho^2\ll 1$, by simply estimating the temperature at $\l =1$, and we get,
\bea
T_{{\rm min}}  =  {\tau (1) \nu (1)^3 \over 4 \pi} { B^2 \over \rho}
\eea
which for the supersymmetric value $k=2/\sqrt{3}$ gives approximately
$T_{{\rm min}} = 2.0932 B^2/\rho$. Of course, at $\lambda=1$ our perturbative analysis breaks down, and so higher order terms could well invalidate this result.
\item If $\tau(1)<0$, then $T=0$ is attained for $\l =\l _c<1$, and the geometry
must have a naked  singularity whenever $\l _c< \l \leq 1$, as it would correspond to negative temperature.
\end{itemize}

\section{Numerical Analysis}
\setcounter{equation}{0}
\label{numerics}

\subsection{Setup}

We turn now to a discussion of our results obtained by numerical integration of the
equations of motion.  The first step is to specify our coordinate system.  We impose the
conditions corresponding to the horizon frame described in section (\ref{parhor}), including choosing $r_+=1$ and
$U'(1)=1$.    This coordinate system will inevitably break down in the limit of vanishing
temperature, since in that case we would have $U'(r_+)=0$, and no rescaling of the
time coordinate can bring  us to our chosen gauge.  We will see this breakdown
occurring explicitly in the numerics.

Solutions in this gauge are parameterized by the values of $b$ and $q$, both of which
we take to be non-negative without loss of generality.  Choosing a value for the pair $(b,q)$
fixes initial data at the horizon, and then we can integrate out to the asymptotically
AdS$_5$ region at large $r$ (we used Maple to do this).

From the large $r$ form of the
obtained solution we can then compute the numerical coefficients ($v$, $w$, etc.) appearing in
(\ref{ub}).  We then convert these into physical quantities using the formulas given in
section \ref{physquant}. The expressions we will be using in the following are
\bea
B&=& {b\over v} \\ \no
T&=& {\gamma_c \over 4\pi} \\ \no
s &=& {1\over 4\sqrt{v^2 w\gamma_c^2}} \\ \no
\rho &=& \gamma_c(e_3 - c_0 p_3)
\eea
with  $\gamma_c=1/\sqrt{1-wc_0^2}$.

It is most illuminating to provide plots of entropy density versus temperature with the
magnetic field and charge density held fixed.  However, it only makes sense to keep fixed
the dimensionless ratio $B^3/\rho^2$.  Similarly, it is only meaningful to compute dimensionless
versions of the entropy and temperature, and for these we choose
\bea
{s \over (\rho^2 +B^3 )^{1/2}}~,\quad\quad   {T\over (\rho^2 +B^3)^{1/6} }
\eea

An instructive special case is the Reissner-Nordstrom solution with $B=0$ reviewed in section
\ref{RNsol}.
The solution is originally given in terms of $r_+$ and $\rho$.  Transforming this solution into the gauge used here, we find $q=\rho/r_+^3$, along with
\bea\label{RNcurve}
{s \over \rho}={1\over 4q} ~,\quad\quad   {T\over \rho^{1/3} } = {1\over 4\pi}\left({4-{2\over 3}q^2\over q^{1/3} }\right)
\eea
Note that the extremal limit in this parametrization is $q=\sqrt{6}$ with
$s/\rho = 1/(4\sqrt{6}) \approx  .102$.   A consistency check on our numerics is that
we recover the curve described by (\ref{RNcurve}) along with the extremal endpoint.

The next step is to determine the region of the $(b,q)$ parameter space that gives rise
to smooth solutions.  The boundary of this region depends on the value of $k$.
Numerical integration shows that as we move out radially from the origin in the $(b,q)$   plane we eventually find that some of the parameters $v$, $w$, etc. start to
diverge or go to zero  as we approach a curve in the $(b,q)$ plane, depicted by the red lines in Fig. \ref{flow}.  
The analytic form of this curve
is only known at $k=1$, where it is given by $q^2 +2b^2=6$.    Points on the $k=1$ critical curve correspond
to extremal warped AdS$_3$ black holes, as was described in  section (\ref{warped}).
For the other values of $k$ that
we consider, the curve acquires some bulges, but continues to look roughly like that for $k=1$.
In the generic case, we determined the critical curve numerically by evolving outward
until $\gamma_c$ exceeds some specified value, which we take to be roughly $12$.  This
requires fine tuning $(b,q)$ to the critical curve to roughly four decimal places.
The $k=0$ case is special since here $\gamma_c=1$ exactly for all solutions; here we
locate the critical curve by looking for a divergence in $1/v$.  In all cases, the
dimensionless temperature tends to zero as we approach the critical curve.

Once the critical curve has been identified, we can set up a grid in the $(b,q)$ plane, and
scan over the gridpoints.  We took about $12,000$ roughly evenly spaced gridpoints.
Once data at the gridpoints has been obtained we can search for curves along which
$B^3/\rho^2$ is approximately constant.  These curves were illustrated schematically by the flow diagrams in  Fig. \ref{flow}. Finally, we plot the entropy versus temperature
for points along such a curve.

\subsection{Results}

We will discuss three cases: $k=0$, $k=1$, and $k=2/\sqrt{3}$, the latter being the supersymmetric value.    We expect that these
are representative of the general cases $k<0$, $k=1$ and $k>1$.

\begin{figure}[h]
\begin{centering}
\includegraphics[scale=0.65]{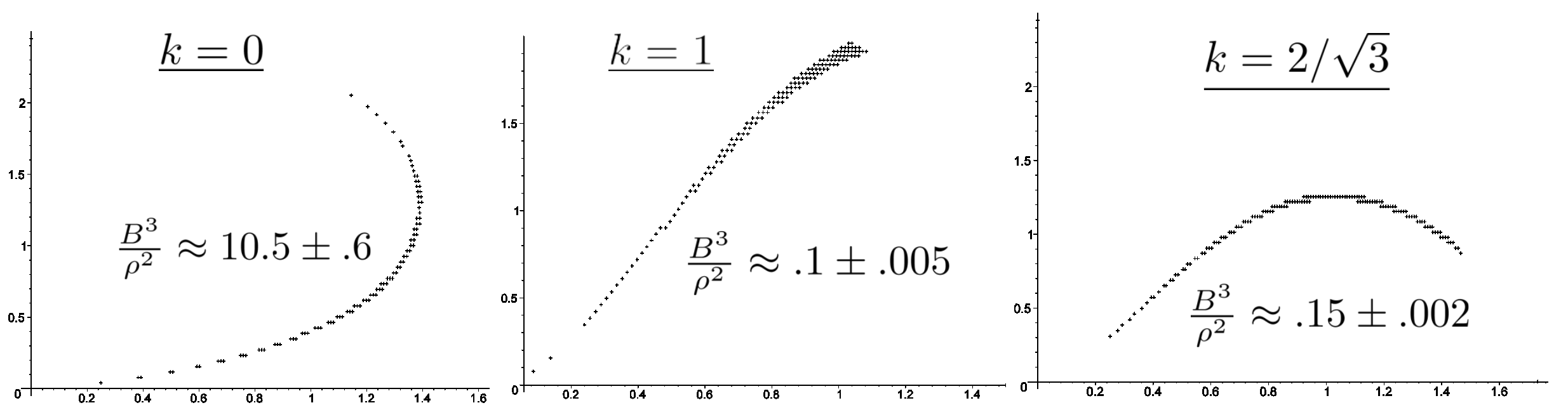}
\caption{Points in the $(b,q)$ plane ($b$ on horizontal axis) obtained by scanning over
a grid and keeping points with $B^3/\rho^2$ fixed within some interval.  These numerical
plots are to be compared with the schematic flows illustrated in Fig. \ref{flow} }
\label{combined.flow}
\end{centering}
\end{figure}

Curves of approximately fixed $B^3/\rho^2$ are shown in Fig. \ref{combined.flow}, which
may be compared with the schematic version in Fig. \ref{flow}.
In the $k=0$ and $k=2/\sqrt{3}$ cases, the curves appear to be heading towards $b=0$ and
$q=0$ respectively.   As they do so, they begin to approach very near to the critical curve discussed above.  We cannot follow them all the way there, as our limited precision prevents
us from collecting data points arbitrarily close to the critical curves.   It is conceivable
that the true curves instead terminate at some location on the critical curves, as apparently
occurs in the $k=1$ case.

\begin{figure}[h]
\begin{centering}
\includegraphics[scale=0.75]{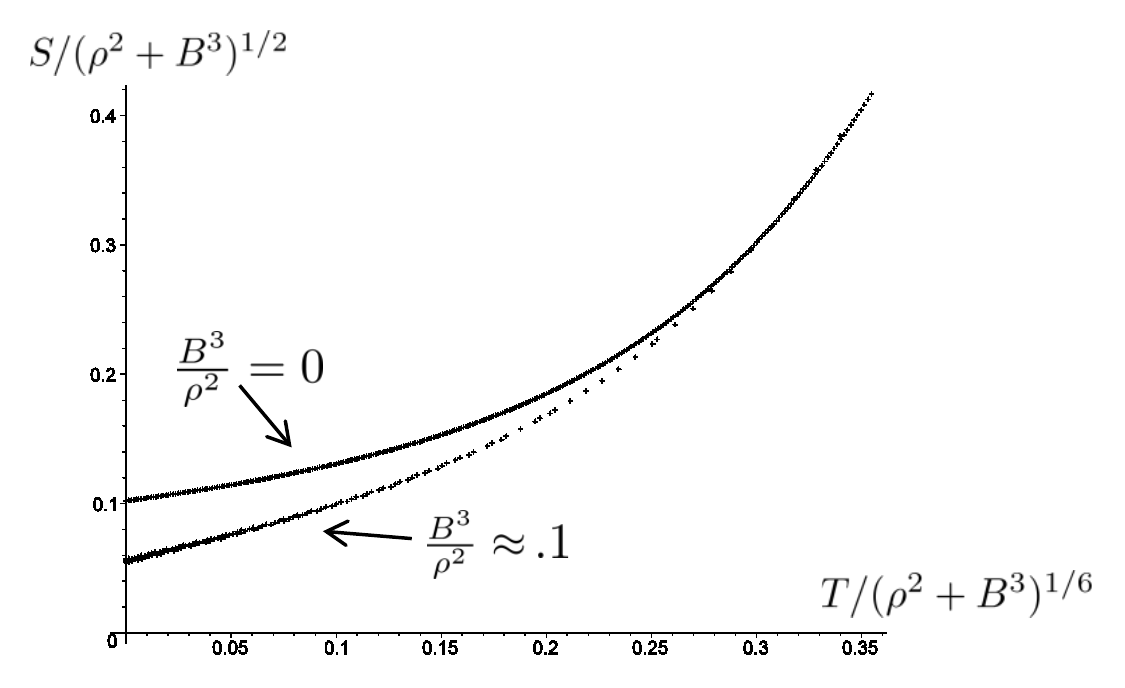}
\caption{Entropy versus temperature for $k=1$ and $B^3/\rho^2 \approx .1 \pm .005$ }
\label{entropy.k.1}
\end{centering}
\end{figure}

Given the points along a fixed $B^3/\rho^2$ curve, we can construct a plot of entropy
versus temperature.  Such plots for the  $k=2/\sqrt{3}$ and  $k=0$  cases are displayed
in Figs. \ref{ksusyentropy} and \ref{entropy.k.0}.  In Fig. \ref{entropy.k.1} we show
the corresponding plot for $k=1$.  
In all of these plots we compare the finite $B$ results against those for $B=0$.  The
$B=0$ curves represent the Reissner-Nordstrom black brane solution, and reproduce
numerically the curve described in (\ref{RNcurve}).

In the $k=0,1$ cases, the entropy appears to go to a finite value at extremality, even though a
singularity seems to be developing at the horizon in this limit, as was discussed in section
\ref{interp}.   By adjusting $B^3/\rho^2$, we can tune this limiting entropy to any desired
value between $0$ and that of the Reissner-Nordstrom solution at $B=0$.  In the plots
we  have chosen the  value of $B^3/\rho^2$ such that the extremal entropy is roughly half of
its maximal value.  We see that this requires a much larger magnetic field in the $k=0$ case
as compared to $k=1$.  At $k=1$ the entropy is controlled by the near-horizon warped AdS$_3$
black hole solution.  

The situation for $k>1$ appears to be very different, at least for the values that we have
studied.   The main effect, which becomes more pronounced at larger $k$, is that the entropy
decreases substantially at low temperatures, and appears to be headed towards zero, until our
numerics break down.  For the supersymmetric case $k=2/\sqrt{3}$ shown in 
Fig. \ref{ksusyentropy} the effect is relatively modest due to the fact that $2/\sqrt{3}\approx 1.15$ is not too much larger than $1$.  But even in this case it is evident that at low
temperature a value of $B^3/\rho^2$ much less than $1$ causes a decrease in the entropy by 
factor much larger than $1$, and the effect grows as the temperature is decreased. 

The main property driving this behavior is the apparent location of the endpoint of the
flows at fixed $B^3/\rho^2$.   We can locate the endpoint numerically by setting $q=0$ and
increasing $b$ until a singularity (or numerical breakdown) occurs.   Calling this endpoint
value $b_c$, we find  $b_c=\sqrt{3}=1.732$ for $k\leq 1$,  but  $b_c < \sqrt{3}$ for $k>1$.
For $k=2/\sqrt{3}$ we find $b_c \approx 1.568$.  On the other hand, with $q=0$  we showed that only for $b=\sqrt{3}$ does a smooth, extremal, near-horizon geometry exist.  Since there
is no candidate smooth endpoint, it seems likely from this perspective
that the  flows terminate in a singularity at the point $(q=0,b=b_c)$.  The metric functions we compute certainly
behave very badly as this point is approached, but it is difficult to completely disentangle
physical divergences from a breakdown of our numerics.  

The breakdown of our numerics at very low temperature is a reflection of our gauge fixing choice $U'(r_+)=1$.   Because of this, we are not able to unambiguously determine whether the $k>1$ 
curves end at zero temperature and entropy, or terminate before then in a singularity.  
But we certainly do not see any evidence of a finite entropy endpoint at zero temperature, 
a conclusion which is bolstered both by our perturbative analysis and by the non-existence
of any candidate near extremal geometries to match onto.   The most likely scenario seems
to be that finite entropy at extremality requires the magnetic field to be fine tuned to zero. 
  One clear goal for the future is to  improve the numerical treatment to allow a more refined
study of the very low temperature regime.

\section{Discussion}

In this work we have constructed asymptotically AdS$_5$ black brane solutions carrying
nonzero charge density and magnetic field.   They were found analytically in a perturbative
expansion for small $B$, and numerically for general values.   The most interesting results
centered around the low temperature regime, where we found a sensitive dependence on both
the magnetic field and the value of the Chern-Simons coupling $k$.  For $k>1$, including
the supersymmetric value of $k=2/\sqrt{3}$, we found that a small magnetic field causes
a rapid decrease of the entropy at low temperatures.  We proposed an AdS/CFT version 
of Nernst's theorem and the third law of thermodynamics consistent with the observed behavior.
More general tests were left to the future \cite{inprogress}.

\sm

$k=1$ emerged as a special value, for here, and only here, there exist smooth finite entropy
extremal near-horizon geometries carrying charge and magnetic field.   We identified these,
and their nonextremal generalizations, with one class of warped AdS$_3 \times R^2$ black hole
solutions studied in \cite{Anninos:2008fx}  (without the $R^2$ factor) in the context of 
topologically massive gravity. We found that these could be connected to asymptotically
AdS$_5$ spacetimes, although a singularity in the interpolating solution develops at the horizon in the strict extremal limit.

\sm

It is an intriguing question as to whether the value $k=1$, which appears special
from the point of view of supergravity solutions, has a special significance also on the 
CFT side, perhaps because it corresponds to a special embedding of the gauge field
$U(1)\subset SU(4)_R$.

\sm

There are some other contexts in which sufficiently large values of the Chern-Simons
coupling $k$  causes novel effects.  In the recent work \cite{Nakamura:2009tf} it was found that sufficiently large $k$  can cause an instability in Reissner-Nordstrom black brane solutions.
In \cite{Gauntlett:1998fz} it was found that large $k$ likely causes an instability of spinning
black hole solutions.  In both of  these cases $k$ must be larger than the supersymmetric
value for an instability to occur, while we have seen here that the extremal Reissner-Nordstrom
entropy is apparently destabilized even at the supersymmetric value.  

\sm

There should be many applications of these solutions to the study of condensed matter and 
finite density QCD.  Transport properties can be computed in these backgrounds; although
we lack analytical solutions, a numerical treatment should be  tractable.  

\sm

We conclude with a curious observation concerning the frequent appearance of $3/4$ in this 
subject.  To wit: the ratio of the high temperature entropy in  gravity to that in the
gauge theory is $3/4$; the corresponding ratio of the low temperature entropy in the presence of a magnetic field is $\sqrt{4/3}$; and the ratio of the ``special" value of $k$  ($k=1$ in our conventions) to the supersymmetric value is $\sqrt{3/4}$.   These three factors are not logically related in any obvious way.  Perhaps this is just  coincidence. 

\newpage

\noindent { {\large \bf Note Added:}} 
\medskip

Relaxing the gauge condition $P(r_+)=0$ by treating $C'(r_+)$ as a 
tunable free parameter, and using numerical analysis with much higher 
precision than was used in the present paper, it has become possible
to explore lower magnetic fields and lower temperatures in a
reliable manner \cite{D'Hoker:2010rz}. The results confirm that,
for the supersymmetric value $k=2/\sqrt{3}$, the entropy density
vanishes linearly with $T$ when $B^3/\rho^2 > 0.124569$, but reveals 
that for magnetic fields smaller than this value, the entropy density 
does {\sl not} vanish at $T=0$. It is established in  \cite{D'Hoker:2010rz}
 that the point $T=0$, $B^3/\rho^2 \sim 0.124569$
corresponds to a quantum critical point with dynamical scaling 
exponent $z=3$. For any fixed $B^3/\rho^2$ above the critical 
value $0.12459$, and possibly for all values, the improved numerics 
also indicate that the flows towards zero temperature  
end at the AdS$_3\times R^2$ fixed point at $b=\sqrt{3}, q=0$,
corresponding to the dot in Figure 2. It is expected that the low
magnetic field phase should ultimately be unstable against turning 
on further couplings $\lambda _i$. 

\bigskip\bigskip

\noindent { {\Large \bf Acknowledgments}}

\bigskip

\noindent  
We thank Vijay Balasubramanian, Geoffrey Compere, Jan de Boer, Frederik Denef, Stephane Detournay, Jerome Gauntlett, Finn Larsen, Alex Maloney, and Joan Simon for helpful discussions
and correspondence.

\appendix

\section{Asymptotic boost symmetry}
\label{AsyBoost}

The Ansatz  (\ref{anz1}), (\ref{anz2}) is  covariant under boosts in the direction
of the magnetic field. Assuming the metric $ds^2$ to be asymptotically
$AdS_5$, the boundary space-time coordinates $x_\mu$ may be rescaled
so that its asymptotic behavior, as  $r \to \infty$, is given by,
\bea
\label{asym1}
ds^2 \sim {dr^2 \over r^2} + r^2 \left ( - dt^2 + dx_1^2 + dx_2^2 +dx_3^2 \right )
\eea
Performing a boost in the $x_3$-direction
\bea
\label{boost}
t & = & \g_c (\ti t + \b \ti x_3) \hskip 1in \g_c^2(1-\b^2)=1
\no \\
x_3 & = & \g_c (\ti x_3 +\b \ti t) \hskip 1.1in |\b|<1
\eea
produces a field strength and a metric of the same form as (\ref{anz1}) and
(\ref{anz2}), but with the coordinates
$r,t,x_3$ replaced by $\ti r, \ti t, \ti x_3$ and the functions $E,P,U,V,W,C$ of $r$
replaced by the functions $\ti E, \ti P, \ti U, \ti V, \ti W, \ti C$ of $\ti r$ respectively, leaving
$x_1, x_2$, and $B$ unchanged. (A reparametrization $ r \to \ti r$ is generally needed
to put the $U$-function back into the gauge of the Ansatz.)
The relation between the transformed and original Maxwell fields are as follows,
\bea
\label{boost1}
\ti E  d \ti r & = & \g_c \Big ( E - \b  P \Big ) dr
\no \\
\ti P d \ti r & = & \g_c \Big ( P - \b  E \Big ) dr
\eea
while for  the metric fields, we have $\ti U ^{-1} d \ti r = U^{-1} dr $, and
$\ti V  = V$, as well as,
\bea
\label{boost2}
\ti U  - e^{2\ti W } \ti C^2
& = & \g_c^2 U - \g_c^2 \Big (C + \b  \Big )^2 e^{2W}
\no\\
e^{2 \ti W} & = & \g_c^2 \Big (1 + \b C  \Big )^2 e^{2 W} - \g_c^2 \b^2 U
\eea
and the following transformation law between $\ti C$ and $C$,
\bea
\label{boost3}
\ti C  = {  \Big ( C +\b  \Big ) \Big ( 1 + \b C  \Big ) e^{2 W} -  \b U
\over
 \Big (1 + \b C \Big )^2 e^{2 W}   -  \b^2 U }
\eea
If $C(\infty)=0$, then it follows that $\ti C(\infty)=0$ for all $\b$. At an event horizon,
$r=r_+$, where $U(r_+)=0$, we have a simplified formula,
\bea
\label{boost4}
\ti C( \ti r_+) = {  C(r_+) +\b  \over 1 + \b C(r_+)  }
\eea
As a result, in these coordinates, the value of $C$ at the horizon
characterizes the velocity of the configuration. Performing a boost
$\beta = - C(r_+)$ brings the solution to its rest frame.

\section{Relation between horizon and asymptotic frames}
\setcounter{equation}{0}
\label{AsyCoord}

Our solutions being asymptotically $AdS_5$ implies that by
rescaling $t,x_1, x_2, x_3$, and performing an $\a$-transformation
and a boost, the asymptotic behavior of the field strength and of the metric
as $r \to \infty$  may be put in standard form,
\bea
\label{bnd1}
F & = & E dr \wedge d \ti t + B d \ti x_1 \wedge d \ti x_2 + Pd \ti x_3 \wedge dr
\no \\
ds^2 & \sim & {dr^2 \over r^2}
+ r^2 \left ( - d\ti t^2 + d \ti x_1^2 + d \ti x_2^2 +d \ti x_3^2 \right )
\eea
In general, the coordinates $\ti t, \ti x_i$ will be different from the
coordinates $t,x_i$ used in specifying initial conditions at the horizon.
In the coordinates $t,x_i$, the asymptotics of the functions take the form displayed in
(\ref{ub}) and (\ref{ubb}).
The rescaling and $\a$-transformation relating the coordinates
$t,x_i$ and $\ti t, \ti x_i$ is given by,
\bea
\ti t =  t \hskip 0.7in \ti x_{1,2} = \sqrt{v} x_{1,2}  \hskip 0.7in
\ti x_3 = \sqrt{w} \left (  x_3 + c_0  t \right )
\eea
This is combined with a constant shift in the $r$ coordinate by an amount $r_0$.
Since the magnetic field term in $F$ is $r$-independent, we
obtain a simple relation between the magnetic field $b$
at the horizon, and the magnetic field $B$ at infinity,
\bea
b = v B
\eea
The functions $\ti U, \ti V, \ti W, \ti C, \ti \cE,\ti P$ of the full metric in the
coordinates $\ti t, \ti x_i$ are then given by,
\bea
\label{uc}
\ti U= r^2 + {u_2 \over r^2} -{ 2 B^2 \over 3}  {\ln r \over r^2} +\cdots \hskip 0.15in
& \hskip 1in &
\ti C= {\sqrt{w} \, c_4 \over r^4} +\cdots
\no \\
e^{2 \ti V} = r^2+ {v_2 \over  v r^2} +{B^2 \over 3}  {\ln r \over r^2} +\cdots \hskip 0.1in
&&
\ti \cE = {e_3 \over r^3}+\cdots
\no \\
e^{2 \ti W} = r^2- {2 v_2 \over v r^2} -{ 2B^2 \over 3}  {\ln r\over r^2} +\cdots
&&
\ti P= {p_3/\sqrt{w}  \over r^3}+\cdots
\eea
While the metric at the boundary of $AdS_5$ now takes the standard
form of (\ref{bnd1}), the metric $ds_H^2$ at the horizon has been rescaled
and $\a$-transformed to become,
\bea
ds_H^2 = {1 \over v} (d \ti x_1^2 + d \ti x_2^2) + { 1 \over w} (d\ti x_3 - c_0 \sqrt{w} d \ti t)^2
\eea
corresponding to a solution moving with velocity $c_0 \sqrt{w} $.

\sm

We can move to the rest frame of the solution by performing a boost in the
$\ti x_3$-direction, thereby removing the cross term in $ds_H^2$  while preserving
the conformal boundary metric,
\bea
\ti t= \gamma_c(\hat t+\sqrt{w}c_0 \hat x_3)~,\quad\quad
\ti x_3 = \gamma_c (\hat x_3+ \sqrt{w}c_0 \hat t)~,\quad\quad
\gamma_c={1\over \sqrt{1-w c_0^2}}
\eea
In these coordinates the field strength is
\bea
F= \gamma_c E d\hat r \wedge d \hat t
+ {b\over v} d\hat x_1 \wedge d \hat x_2 +
\left ( {P\over \sqrt{w}\gamma_c}  -\sqrt{w} c_0 \gamma_c E \right )
d\hat x_3 \wedge d \hat r
\eea
and the metric on the horizon is
\bea
ds_H^2 =   {1\over v}({d \hat x_1}^2 +{d \hat x_2}^2)
+{1\over w\gamma_c^2 } {d\hat x_3}^2
\eea
which corresponds to a solution at rest in the hatted frame. This hatted (asymptotic) frame is the one that we use to express physical quantities in the CFT. 

\subsection{Evaluating entropy, temperature and chemical potential}

Now we can give expressions for the various physical quantities in these coordinates.
The entropy density $s$ is read off from metric on the horizon, and is given by
\bea
s = {1\over 4} {1\over \sqrt{v^2 w \gamma_c^2} }
\eea
The factors in the denominator ensure that $s$ is the  physical entropy density per unit volume, as measured in the CFT.

To evaluate the temperature, we first recall its behavior under a boost.  Let $T_0$ be the temperature of some system  in the rest frame.   Then in a boosted frame the temperature is
given by $T=  T_0/\gamma_c$, as can be seen by writing the Boltzmann factor $e^{-E/T_0}$ in
terms of the boosted energy and momentum.    Now in the original (unhatted) frame
the temperature is  $\tilde{T}=U'(r_+)/(4\pi)$.   This
represents the moving frame, while the rest frame is one in which
$C(r_+)=0$, which is the hatted frame.   Therefore, the temperature in the rest frame, which
is what we're interested in, is
\bea
\hat T=T_0 = \gamma_c \ti T = {\gamma_c U'(r_+)\over 4\pi}
\eea
From now on, when we write $T$ we have in mind $\hat{T}$.

We also want the chemical potential, given by the asymptotic value of $\hat A_t$.
Since we set $\hat A_t=0$ at the horizon we have
$ \hat A_t |_\infty = \int_{r_+}^\infty \! d\hat{r}~  \hat F_{ rt}$.
This gives
\bea
\label{uj}
\mu = \hat A_t \Big |_\infty =  \gamma_c \int_{r_+}^\infty \! dr ~E
\eea
%
%
If $kB\not=0$, we may use the second Maxwell equation, in the form
given in (\ref{max1a}) to recast $E$ as a total $r$-derivative. The chemical
potential may now be evaluated in the original unprimed coordinates,
where the magnetic field is $b$, and we find,
\bea
\label{uja}
\mu = {3  \gamma_c v \over 8 kb }
\left(  \sqrt{w} c_0 e_3 - {p_3 \over \sqrt{w}}\right)
\eea

\subsection{Evaluating the Maxwell current}

We identify the Maxwell current in terms of the on-shell variation of the
Maxwell-Einstein action of (\ref{action1}), with respect to the Maxwell field
at the boundary of $AdS_5$,
\bea
\label{yd}
\delta S = \int\! d^4x\sqrt{-\gamma^{(0)}}J^\mu \delta A_\mu
\eea
Here, $\gamma^{(0)} _{\mu \nu}$ is the asymptotic conformal metric defined
by $\gamma_{\mu \nu}/r^2 \to  \gamma^{(0)}_{\mu \nu}$ in the limit $r \to \infty$, where
$\gamma_{\mu\nu}$ is the metric induced from $g_{\mu\nu}$ on a surface of constant $r$. 
The on-shell variation of the action with respect to the gauge field may be calculated
from (\ref{action1}) and we find the following expression for the current,
\bea
\label{ye}
4\pi G_5J^\mu = -\Big(r^3 F^{r\mu}+r^4 (\ln r) \p_\nu
F^{\nu\mu} +{k\over 3} \eps^{\alpha\beta\gamma\mu}A_\alpha
F_{\beta\gamma} \Big)
\eea
Here, $\eps^{\alpha\beta\gamma\mu}$ denotes the volume form for $\gamma^{(0)}_{\mu\nu}$.
Using the leading forms $g^{rr}=r^2$ and $g^{\mu\nu}=\gamma_{(0)}^{\mu\nu}/r^2$ we can
write this as
\bea
\label{yf}
4\pi G_5 J^\mu = 
\Big(r^3 \gamma_{(0)}^{\mu\nu}F_{r\nu}  + (\ln r)
\gamma_{(0)}^{\mu\alpha}\gamma_{(0)}^{\nu\beta} \p_\nu F_{\alpha\beta} +{k\over 3}
\eps^{\alpha\beta\gamma\mu}A_\alpha F_{\beta\gamma} \Big)
\eea
Since we only consider solutions with constant field strength on the
boundary the middle term will not contribute.

\subsection{Evaluating the current in the rest frame}

The current can be computed directly in the double primed coordinates system
as
\bea
\label{yfax}
{ 4\pi G_5} \hat J^t &=&
\left ( \hat e_3-{2 k \hat B \over 3 \sqrt{\hat v^2 \hat w}} \hat A_3 \Big |_\infty \right )
\\ \no
{ 4\pi G_5}\hat J^{1,2}&=&0
\\ \no
{ 4\pi G_5}\hat J^3 &=&
\left ( {\hat p_3 \over \hat w}+{2k\hat B \over 3\sqrt{\hat v^2 \hat w}} \hat A_t \Big |_\infty
\right )
\eea
The double primed coordinates have been defined so that $\hat v= \hat w=1$.
From the formulas given above we also have,
\bea
\hat B = {b\over v}~,\quad\quad
\hat e_3 = \gamma_c(e_3-c_0 p_3)~,\quad\quad
\hat p_3 = \gamma_c \left({p_3 \over \sqrt{w}}-\sqrt{w}c_0 e_3 \right)
\eea
and
\bea
\hat A_t \Big |_\infty =
-{3 \gamma_c v \over 8 kb} \left ( {p_3 \over \sqrt{w}}-\sqrt{w}c_0 e_3 \right )
=-{3 v\over 8kb} \hat p_3
\eea
We therefore finally have 
\bea
4\pi G_5 \hat J^t
&=&\gamma_c(e_3-c_0 p_3)  -{2kb \over 3v} \hat A_3 \Big |_\infty
\\ \no
4\pi G_5 \hat J^3&=& {3\over 4}\gamma_c \left({p_3 \over \sqrt{w}} -\sqrt{w}c_0 e_3\right)
\eea
We note  that $\hat A_3 \Big |_\infty$ can be chosen arbitrarily as long as $g_{33}$ is finite at the horizon (though we should set it
to zero unless we want to add a source for $J^3$ in the CFT partition function),
while $\hat A_t \Big |_\infty$ is given in (\ref{uj}).

\section{Factorized Solutions}
\setcounter{equation}{0}
\label{factorized}

In the general 5-dimensional solutions we have obtained here, the 2-dimensional
$x_1,x_2$-plane perpendicular to the magnetic field is warped over the remaining
3-dimensional space. The near-horizon geometry of these solutions, however,
invariably reduces to a space-time in which the $x_1,x_2$-plane factorizes from
the solutions because the field $V$ becomes $r$-independent. This raises the
question as to the structure of the most general factorized solution. In this appendix
we shall show that, modulo certain regularity conditions, the only factorized
solutions are of the type given in section 4.

\sm

We begin by proving the following auxiliary result:  if the function $P$ vanishes,
then we only have the factorized $k^2=1$ solution of section 4.4. Next,
assuming now that $P\not=0$, but $V=0$, we show that again only the
factorized $k^2=1$ solution of section 4.4 exists.

\subsection{$P=0$ leads to the near-horizon geometry}

By shifting $V,W$ by constants, we may always assume that $V(r_+)=W(r_+)=0$.
Recall that the magnetic field in these coordinates is denoted by $b$.
We assume $kb\not=0$ and $q\not=0$.  We begin by showing that
the condition $P=0$ leads to $U'',V,W, \cE$ constant. When $P=0$,
equations M1, M2, E1 may be integrated, and give,
\bea
\cE & = & q \, e^{-2V-W}
\no \\
C' & = & 2 kb \, e^{-2V-W}
\no \\
C' & = & 2kb \, e^{-2V-3W}
\eea
Comparing the two expressions above for $C'$, we find that
$W=0$ identically. Taking the difference between E2 and E3 (for $W=0$),
we find,
\bea
0 = 24 - 4 \cE^2 + 4 b^2 e^{-4V} - 3 (C')^2
\eea
Substituting the above expression for $C'$ and $\cE$, we find,
\bea
0 = 24 + e^{-4V} \left ( - 4 q^2 + 4 b^2 -12 k^2 b^2 \right )
\eea
Since the terms in the parenthesis are constants,  $V$
must be constant, and hence $V=0$ in view of the boundary condition.
Thus, $\cE$ and $C'$ are constant. Equation E2 now requires
$0 = 2 k^2 b^2 - 2 b^2$ which requires $k^2=1$. Equation E3 requires
\bea
0 = 6 - q^2 - 2 b^2
\eea
which is precisely the boundary curve equation for $k^2=1$. CON is automatic, and
E4 gives $U'' = 24$. But this gives precisely the factorized solution of section 4.4.

\subsection{$V=0$ leads to the near-horizon geometry}

Next, we assume $P\not=0$ and $V=0$, so that the geometry
is factorized. Under this assumption, equation M2 may be traded for the
constraint equation, since upon differentiation, CON will require the use
of M2 when $P \not=0$.  Clearly, we must now retain E4 as an
independent equation.
Equations E2, E3, and CON are equivalent to the following equations,
\bea
\label{reda}
\cE^2 & = & 6 - 2b^2 + UP^2 e^{-2W}
\no \\
U'W' & = & - \half (C')^2 e^{2W} + 2 b^2
\no \\
- W'' -(W')^2 & = & 2 P^2 e^{-2W}
\eea
By eliminating $\cE^2$, $(C')^2$, and $P^2$ from E4, using the above
equations, we obtain a relatively simple equation relating only $U$ and $W$,
\bea
\label{redU}
U'' + 3 U'W' + 2 U W'' + 2 U (W')^2 =24
\eea
Finally, equation M1 may be viewed as giving $P$ in terms of $\cE$ and $W$,
\bea
\label{redP}
P = -{1 \over 2kb} \left ( \cE e^W \right )'
\eea
and E1 may be integrated upon using M1, to give,
\bea
\label{redC}
C' = - {1 \over kb} \cE^2 e^{-W} + \a e^{-3W}
\eea
where $\a$ is an integration constant. As promised, we omit equation M2.

\sm

Equation (\ref{redU}) may be readily integrated in terms of the function $\tau$,
defined by,
\bea
\tau (r) \equiv \int ^r _{r_0} dr' \, e^{W(r')}
\eea
and we have
\bea
U= { 12 \over (\tau')^2 } (\tau^2 - \tau_0^2)
\eea
where $r_0$ and $\tau_0$ are integration constants. Using the relations
$\tau ' = e^W$ and,
\bea
W' = { \tau '' \over \tau'} \hskip 1in
W'' + (W')^2 = { \tau ''' \over \tau '}
\eea
and eliminating $\cE, U, P,W$ from (\ref{reda}) in favor of $\tau$ and
$\hE = \tau ' \cE$, we find the following three equivalent equations,
\bea
\label{redE}
\hE^2 & = & (6 - 2b^2)(\tau')^2  - { 6 \tau ''' \over \tau' } (\tau^2 - \tau_0^2)
\\
(\hE')^2 & = & - 2 k^2 b^2 \tau ' \tau '''
\no \\
0 & = &  48 (\tau'')^2 (\tau^2-\tau_0^2) - 48 \tau (\tau')^2 \tau ''
+ 4 b^2 (\tau')^4 - {1 \over k^2 b^2} \left ( \hE^2 - \a kb \right )^2
\no
\eea
Finally, $\hE$ and $\hE'$ may be eliminated from the above
equations to yield two equations for $\tau$. Eliminating $\hE^2$
between the first and third equations yields an equation of third order
in $\tau$. Eliminating $\hE'$ requires some extra care, as we do not
wish to unduly increase the order of the resulting equation, or introduce
square roots. To obtain a second equation of third order, we take the
product of the first two equations giving $[(\hE^2)']^2/4$,
and  use the last equation to eliminate $\hE^2$ from this.
The results are as follows,
\bea
\label{redtau}
0 & = & 6 (6-2b^2) \tau^2 (\tau ''')^2 + 12 \tau_0^2 b^2 (\tau''')^2
+ (6-2b^2)^2 (\tau')^2 (\tau'')^2
\no \\ &&
- 12 (6-2b^2) \tau \tau ' \tau '' \tau '''
+ 2 b^2 (6-2b^2) (\tau')^3 \tau'''
\no \\ && \no \\
0 & = & 48 k^2 b^2 (\tau'')^2 (\tau^2- \tau_0^2) - 48 k^2 b^2 \tau (\tau')^2 \tau ''
+ 4 k^2 b^4 (\tau')^4
\no \\ &&
 -  \left ( (6-2b^2) (\tau')^2 - 6 { \tau ''' \over \tau'} (\tau^2 - \tau_0^2)
 - q^2 - 2 k^2b^2 \right )^2
\eea
We shall now examine the existence of joint solutions to both equations,
which solve the boundary conditions, $U(r_+)=W(r_+)=P(r_+)=0$, and
$C'(r_+) = 2kb$. These conditions translate as follows
\bea
\tau (r_+) & = & \tau _0 \hskip 1in  kb \a = q^2 + 2 k^2 b^2
\no \\
\tau' (r_+) & = & 1
\no \\
\tau '''(r_+) & = & 0 \hskip 1in \hE'(r_+)=0
\eea
Evaluating the first equation of (\ref{redE}) at the horizon, using the above
boundary conditions, gives the boundary curve relation,
\bea
\label{fam}
q^2 + 2 B^2 = 6
\eea
Evaluating both equations of (\ref{redtau}) at the horizon, and using the boundary conditions, $\tau'(r_+) =1$ and $\tau'''(r_+)=0$, we find,
\bea
\label{redp}
0 & = & (6-2b^2) \tau''(r_+)^2
\no \\
0 & = & - 48 k^2 b^2 \tau_0 \tau''(r_+) + 4 k^2 b^4 - 4 k^4 b^4
\eea
where in the last equation we have used the relation (\ref{fam}).
Since we assume that $q \not= 0$, relation (\ref{fam}) gives $6 - 2 b^2 \not= 0$,
and thus we must have $\tau '' (r_+)=0$ from the first equation in (\ref{redp}).
Using this result in the second equation gives altogether
\bea
\tau ''(r_+)=0 \hskip 1in k^2=1
\eea
Note that we have now, in principle,  overdetermined the boundary conditions
even just on a single equation, as we have $\tau(r_+)=\tau_0, \tau'(r_+)=1,
\tau''(r_+)=\tau'''(r_+)=0$ for a differential equation which is of third order.
To proceed further, it appears necessary to make some assumption on the
regularity properties of $\tau$ near the horizon. A general Ansatz consistent
with the above boundary conditions is as follows,
\bea
\label{astau}
\tau (r) = \tau_0 + (r-r_+) +\tau_p  (r-r_+)^p + {\rm higher~orders}
\eea
for any real number $p >3$. Substituting this Ansatz into the first equation of (\ref{redtau})
shows that, as $r \to r_+$, the last term is of order $(r-r_+)^{(p-3)}$ and dominates
the other 4 terms, which all vanish faster as $r \to r_+$. Thus, we must have
$\tau_p=0$, for any $p >3$. Thus, within the class of asymptotic behaviors
given by (\ref{astau}), the expression
\bea
\tau (r) = \tau_0 + (r-r_+)
\eea
is the only solution satisfying the boundary conditions. Clearly, for this
solution, we have $U''= 24$, and $W$ constant, which is the solution of
section 4.4.

\section{Interpolating between extremal BTZ $\times R^2$ and AdS$_5$}
\setcounter{equation}{0}
\label{BTZ}

In section \ref{interp} we alluded to the existence of solutions that interpolate
between a near-horizon extremal BTZ $\times R^2$ geometry and AdS$_5$.  In this
appendix we give the details of their construction.  They can be thought of as infinitely boosted versions of the solutions studied in \cite{D'Hoker:2009mm}.

\subsection{Extremal solutions with momentum}
\label{exBTZ}

In \cite{D'Hoker:2009mm} we found zero temperature solutions interpolating between 
AdS$_3 \times R^2$  (with magnetic flux on the $R^2$) and AdS$_5$.  The metric and field
strength are
\bea
ds^2&=& e^{-2W(r)}dr^2 +e^{2W(r)}(-dt^2 + dx_3^2) +e^{2V(r)}(dx_1^2+dx_2^2) \\ \no
F&=& B dx_1 \wedge dx_2 
\eea
The Einstein-Maxwell equations reduce to
\bea
\label{zerotemp}
&&2V''+W''+2(V')^2 +(W')^2 =0 \\ \no
&& (V')^2 +(W')^2 +4V'W'=6e^{-2W}-e^{-4V-2W}B^2 
\eea
One can (numerically) find solutions with small $r$ behavior $e^{2V}=B/\sqrt{3}$, 
$e^{2W}=r^2$, and large $r$ behavior $e^{2V}= vr^2$, $e^{2W}=r^2$, with $v\approx 1.87$.  
Since $B$ can be set to unity be a coordinate rescaling, there is actually a unique such 
solution.  The solutions are not entirely smooth, as $V$ develops a subleading small $r$
dependence $r^\alpha$ with $0<\alpha <1$, as was discussed in section \ref{interp}.

Physically, these solutions represent the RG flow of ${\cal N}=4 $ SYM theory in the 
presence of a magnetic field.  At low energies the theory is governed by fermions in
the lowest Landau level.  These fermions are free to move  parallel to the magnetic field lines, and give rise to a $1+1$ dimensional CFT at low energies, hence the appearance of AdS$_3$. 
  In \cite{D'Hoker:2009mm} the central charges in gravity and  ${\cal N}=4 $ SYM at vanishing
coupling were compared, and found to differ by a factor of $\sqrt{3/4}$.   Note that the 
theory is nonsupersymmetric even at zero temperature due to the presence of the magnetic field. 

On the CFT side, one should be able to excite one chirality of fermions to arrive at
zero temperature configurations carrying momentum, and we expect corresponding solutions
on the gravity side as well.   The structure of such solutions follows, as in 
\cite{Garfinkle:1990jq,Garfinkle:1992zj}, from the existence of null translational isometries. 

We take as our Ansatz 
\bea
ds^2&=& e^{-2W(r)}dr^2 +e^{2W(r)}(-dt^2 + dx_3^2) +e^{2V(r)}(dx_1^2+dx_2^2)- u(r)(dx_3-dt)^2  \\ \no
F&=& B dx_1 \wedge dx_2 
\eea
with $V$ and $W$ obeying (\ref{zerotemp}).   Plugging in, we find that the Einstein-Maxwell
equations reduce to the following linear equation 
\bea
\label{ra}
u'' +(2V'-W')u'-2\Big(W''+(W')^2+2 V'W'\Big) u =0
\eea

This equation can be solved subject to  the boundary
condition that  $u(r) $ should fall off as $1/r^2$ as $r \to \infty$. The 
solution is given by
\bea
u(r) =  p \, e^{2W(r)} \int _\infty ^{r} d\xi \, e^{-2V(\xi) - 3 W(\xi)}
\eea
where $p$ is an integration constant.  
Given the asymptotics of $V$ and $W$, the precise fall-off is
\bea
u(r) = -{ p \over 4 r^2} + \cdots
\eea

Evaluating the boundary stress tensor on these solutions we find 
\bea
8\pi G_5 T^{tt}&=& -{3\over 2}u_2+{1\over 2}p \\ \no
8\pi G_5 T^{t3}&=& {1\over 2}p
\eea
where $u_2$ determines the large $r$ falloff of $U=e^{2W}$ according to $U=r^2 +{u_2 \over r^2}+\cdots$.    The form of the stress tensor is consistent with a momentum $p$ being carried by chiral excitations.   Another manifestation of this is the formula for the entropy
density, which is
\bea
{1\over G_5} s= 2\pi \sqrt{{c\over 6} L_0}
\eea
with
\bea
L_0 = {T^{t3}\over 2\pi}~,\quad \quad c= {B\over 2G_5}
\eea
The central charge  is the same as was computed in \cite{D'Hoker:2009mm}.

The solutions just described carry momentum but no electric charge or current. 
The latter can be included in a fairly trivial way by turning on constant gauge potentials subject to regularity conditions at the horizon.  According to (\ref{yfa}), the combination
of the Chern-Simons coupling $k$ and the magnetic field $B$ converts such constant 
potentials into charges and currents.  The regularity condition at the horizon is
$A_t+A_3$, and so we can turn on $A_t=-A_3 = A$ for any constant $A$.  This induces
\bea
4\pi G_5 J^t = 4\pi G_5 J^3 ={1\over 2}kBA
\eea

Although we have thus constructed finite entropy, extremal, solutions carrying 
both charge and magnetic field, we see that this requires that we have a nonzero
value of $A_3$ at infinity, which corresponds to a nonzero chemical potential for 
current in the CFT.  If we demand that this chemical potential is zero, corresponding to the boundary condition  $A_3|_{\infty} =0$, then these solutions do not appear.

\end{document}